\documentclass{amsart}
	\usepackage[pdftex]{graphicx}
	\usepackage{siunitx}

	\theoremstyle{definition}
	
	\newtheorem{definition}{Definition}

	\newtheorem{subdefinition}[definition]{Definition}
	
	\newtheorem{property}{Property}
	
	\newtheorem{lemma}{Lemma}	
	\usepackage{algorithm}
	
	\usepackage{algorithmicx}

	\usepackage{algpseudocode}
	\usepackage{tabularx}
	\algrenewcommand\algorithmicindent{0.8em}

	\usepackage{paralist}
	\usepackage{xfrac}

	\usepackage{array}

	\usepackage{url}
	
	\usepackage{calc}
	
\begin{document}

		\title{Neighbor Discovery Latency in BLE-Like Protocols}
		
		\author{Philipp H. Kindt, Marco Saur, Michael Balszun and Samarjit Chakraborty}
		\thanks{All authors are with the Chair of Real-Time Computer Systems (RCS), Technische Universit\"at M\"unchen (TUM), Germany. Email: kindt/saur/balszun/chakraborty[at]rcs.ei.tum.de. \textcopyright 2017 IEEE. DOI: 10.1109/TMC.2017.2737008}
		\maketitle
		
		\begin{abstract}

Neighbor discovery is the procedure using which two wireless devices initiate a first contact. In low power ad-hoc networks, radios are duty-cycled and the latency until a packet meets a reception phase of another device is determined by a random process. Most research considers slotted protocols, in which the points in time for reception are temporally coupled to beacon transmissions. In contrast, many recent protocols, such as ANT/ANT+ and Bluetooth Low Energy (BLE) use a slotless, periodic-interval based scheme for neighbor discovery. Here, one device periodically broadcasts packets, whereas the other device periodically listens to the channel. Both periods are independent from each other and drawn over continuous time. Such protocols provide 3 degrees of freedom (viz., the intervals for advertising and scanning and the duration of each scan phase). Though billions of existing BLE devices rely on these protocols, neither their expected latencies nor beneficial configurations with good latency-duty-cycle relations are known. Parametrizations for the participating devices are usually determined based on a "good guess". In this paper, we for the first time present a mathematical theory which can compute the neighbor discovery latencies for all possible parametrizations. Further, our theory shows that upper bounds on the latency can be guaranteed for all parametrizations, except for a finite number of singularities. Therefore, slotless, periodic interval-based protocols can be used in applications with deterministic latency demands, which have been reserved for slotted protocols until now. Our proposed theory can be used for analyzing the neighbor discovery latencies, for tweaking protocol parameters and for developing new protocols.

\end{abstract}

\section{Introduction}
With the increasing pervasiveness of smartphones and wireless sensors, mobile ad-hoc networks (MANETs), in which all participating devices are battery-powered, have become widely-used. MANETs nowadays are applied e.g., for connecting gadgets to smartphones, for localization beacons, object tracking and mobile hotspots.
In such applications, energy saving is a crucial requirement.

The procedure to establish a first contact between two devices in a MANET is usually referred to as \textit{neighbor discovery}. 
To save energy, the devices usually apply duty-cycling, which means that they repeatedly switch on their radios and sleep in the meantime. A successful discovery can take place if and only if one radio listens on the channel at the same point in time at which another one sends a packet. As the clocks of these devices are completely unsynchronized, the point in time when two devices meet for the first time is random.
A popular example of a protocol that applies such a duty-cycled scheme is Bluetooth Low Energy (BLE) \cite{bleSpec}.
Neighbor discovery is an active area of research and many approaches have been proposed. Mainly, they fall in one of the following categories:\\

\noindent\textbf{Wakeup-Receivers}: Approaches such as \cite{spenza:15}, \cite{magno:16} assume that there is an additional, low power receiver that continuously senses the channel for wakeup beacons. Many wireless systems do not have such an additional receiver. Therefore, they make use of one of the following approaches.
\\
\textbf{Birthday Protocols}: These protocols require every node to switch between a low-power-state, a sending-state and a receiving-state at random points in time with given transition probabilities \cite{mcglynn:02}, \cite{margolies:16}, \cite{chen:16}. Whereas good average latencies can be achieved, the main drawback is the lack of guarantees on upper latency bounds. 
\\
\textbf{Slotted Protocols:} In this protocol family, time is subdivided into fixed-length slots. The most important concept of slotted protocols is that the points in time at which packets are sent (and hence also the number of packets) are always associated to listening phases. In each slot, a device can either sleep or be active. Typically, each device needs to send a packet at the beginning and the end of each active slot and must listen during the entire duration in between \cite{dutta:08}. Multiple patterns of active and sleep-slots have been proposed for providing bounded worst-case latencies. For example, coprimality-based approaches \cite{herman:07}, \cite{dutta:08}, \cite{Kandhalu:10} assume that active slots occur with a certain repetition period. The periods of two devices are chosen such that they are mutually co-prime and hence, the Chinese Remainder Theorem (CRT) guarantees an overlap within a limited number of slots. As an extension, \cite{zhang:12} proposes adding additional active slots to protocols like Disco~\cite{dutta:08} or U-Connect~\cite{Kandhalu:10} for exchanging information on devices discovered by the already known neighbors. 
An alternative scheme is used by quorum-based protocols~\cite{tseng:02}, in which a period of $m$ slots is assumed. The slots of a hyper-period of length $m^2$ are arranged in a $m \times m$-matrix. One node choses a row for determining its active slots, whereas the other one choses a column, such that discovery is guaranteed within a limited number of slots.
Recently, a protocol called \textit{Searchlight} has been proposed in \cite{bakht:12}, where it has been claimed that it outperformed all other quorum- and CRT-based protocols at the time of its publication. In each period of Searchlight, there are two active slots. The first one has a fixed position in each period, whereas the other one repeatedly changes its position in a cyclic manner, thereby deterministically meeting the remote node's fixed slot within a finite number of slots. 
Another quorum-based protocol called \textit{Hedis} \cite{chen:15} is claimed to outperform Searchlight in asymmetric cases, in which both devices operate with different duty-cycles.
Moreover, the parametrizable protocol \textit{Hello} \cite{Sun:14} generalizes the already mentioned protocols \textit{Searchlight} \cite{bakht:12}, \textit{Disco} \cite{dutta:08} and \textit{U-Connect} \cite{Kandhalu:10}, which can be described as special cases of it. The main drawbacks of slotted strategies are a) the restricted set of duty-cycles that can be realized, b) redundant slots \cite{Sun:14}, and c) the requirement to send beacons always directly before or after scanning phases, which constrains the design space.

\noindent\textbf{Slotless, periodic interval-based solutions:} While slotted protocols require that beacon transmissions are temporally coupled to reception phases, slotless, periodic interval (PI)-based  solutions break away from this constraint. Packet transmission occurs with an interval $T_a$, while receptions take place for a certain duration $d_s$ once per period $T_s$. The parameters $T_a$, $T_s$ and $d_s$ are drawn over continuous time and can be chosen freely by the protocol designer. Though such schemes can also be used for symmetric two-way discovery, most existing protocols (e.g., BLE and ANT/ANT+) assign different roles to different devices. In particular, they assume that one device only broadcasts advertising packets, whereas the other one only scans.
One of the first periodic interval-based protocols, STEM-B, has been proposed in \cite{schurgers:02}. It works as follows.

\begin{itemize}
	\item One device, called the \textit{initiator} or \textit{advertiser}, broadcasts advertising packets with a fixed interval $T_a$. 
	\item Another device that is called the \textit{target} or \textit{scanner} continuously listens for packets during $d_s$ amounts of time once per interval $T_s$. We refer to $d_s$ as the \textit{scan window}.
	\item If the scanner receives a packet, it sends a response on a different wireless channel. Once the initiator receives this packet, e.g. by listening after each transmission \cite{bleSpec}, both devices are synchronized.  
\end{itemize} 

\noindent\textbf{Popular slotless protocols:} Though most recent research has considered slotted discovery, adaptations of the slotless scheme described above are widely used in practice. The ANT/ANT+\cite{AntSpec:14} protocol, which is used in over 100 million devices \cite{antWebsite:15}, implements this scheme, with the only modification that it performs the complete discovery procedure on one channel \cite{AntChannelSearch:09}. Further, it has been adopted and slightly modified for BLE, in which an advertiser periodically sends up to three packets in a row on different channels \cite{bleSpec}. This period consists of the sum of an interval $T_a$ and a random delay of up to $\SI{10}{\milli s}$. On the remote side, the scanner listens periodically on one of these three channels for a duration of $d_s$, thereby toggling the channel once per interval $T_s$. 

As we will describe in detail in Section \ref{sec:related_work}, even though such protocols are widely used e.g., in billions of BLE devices, their behavior currently cannot be fully analyzed. Except for the trivial case of $T_a < d_s$, no theory for computing their latencies exists. In particular, it is not known whether they can guarantee any upper latency bounds. Moreover, the impact of different parameter valuations for $T_a$, $T_s$ and $d_s$ cannot be studied in a systematical manner. As a result, there is currently no feasible way to choose optimal protocol parameters, e.g., for BLE. However, parameter optimization is extremely important since, as we will show, unfavorable parametrizations risk lying in a hyperbolic peak and therefore cause long mean latencies and high energy consumptions for the connection setup.

\noindent\textbf{Difference from slotted protocols:} The main reason for this lack of understanding is the complex analysis of PI-based, slotless protocols. Whenever two slots in a slotted protocol overlap in time, discovery is guaranteed, regardless of the actual temporal length of the overlap. Therefore, only a finite number of different temporal offsets between the active slots of two devices need to be considered, which makes the analysis simple. In contrast, in slotless, PI-based protocols, there could be an infinite number of initial offsets with different associated latencies. This makes existing models for slotted protocols unsuitable for slotless solutions. The theory presented in this paper shows that the range of possible initial offsets can be subdivided into multiple, variable-length partitions with constant discovery latencies and hence the number of possibilities becomes finite for PI-based protocols, too. Another challenge is finding the borders of these partitions, which is not straightforward.

In this paper, we for the first time present a mathematical model that can compute the mean and worst-case neighbor discovery latencies of periodic interval-based, slotless protocols for all possible parameter values.

\begin{figure}[t]
	\centering
	\includegraphics[width = 1.0\columnwidth]{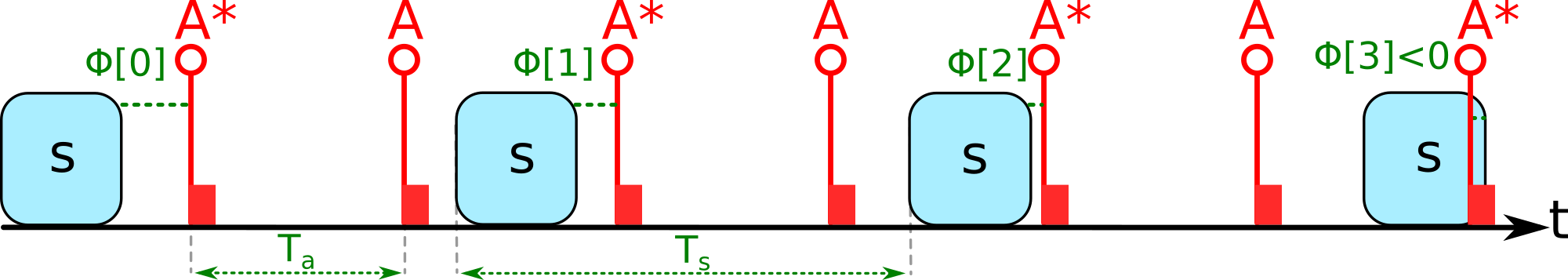}
	\caption{Shrinking distances $\Phi[k]$ between neighboring advertising packets (\textit{A}) and scan windows (\textit{S}). The index $k$ is a counting number for the scan windows.}
	\label{fig:introduction_example} 
\end{figure}

\noindent\textbf{Modeling PI-based protocols:} The key idea of our proposed model is to track the change in the temporal distance between neighboring advertising packets and scan windows over time, rather than considering the corresponding absolute points in time. This temporal distance $\Phi[k]$ is depicted in Figure~\ref{fig:introduction_example} for $k = 0$ to $k=3$. In this example, an advertising packet is the \textit{right neighbor} of a scan window $k$, if it is the closest packet temporally on the right of it. In Figure~\ref{fig:introduction_example}, such neighboring advertising packets are tagged with \textit{A*}, whereas the remaining ones are tagged with \textit{A}. In this example, after each instance $k$ of the scan interval $T_s$, the temporal distance $\Phi[k]$ shrinks by a constant value. This shrinkage occurs repeatedly for multiple subsequent intervals, until the temporal distance falls below the length of one scan window $d_s$. Therefore, a match between the advertiser and the scanner occurs, as for the last scan window of Figure \ref{fig:introduction_example}. In other examples, $\Phi[k]$ grows instead of shrinking, or the distance $\Phi[k]$ shrinks or grows when considering suitable multiples of $T_a$ and $T_s$. Nevertheless, this concept can be generalized to all possible constellations. These generalizations require a more elaborate procedure, which is described in Section \ref{sec:overview}.
By taking into account all possible initial offsets between the first advertising packet and the first scan window, the exact expected value of the discovery latency and its maximum value can be calculated using analytical methods with low computational complexity. 

\noindent\textbf{Comparison with related models:} This technique differs significantly from all known models for PI-based protocols. All existing models are restricted to the trivial case of $T_a < d_s$, and rely on counting the number of $T_a$-intervals until reaching the time-instance at which the scanner switches on its receiver for the first time. Parametrizations with $T_a >  d_s$ are very relevant in practice, since they enable significantly lower duty-cycles of the advertiser and are widely used in the practice, e.g., as a recommended parametrization in the BLE find me profile \cite{BleFindMeProf}. To the best of our knowledge, the technique of tracking the change in the temporal distance among neighboring periodic events, as proposed in this paper, has not been known before, neither for neighbor discovery nor in other domains. This generic technique is suitable for modeling possible overlaps of all kinds of periodically repeated actions, and can therefore be applied in other applications, e.g., for accelerating discrete-event simulations too.

\noindent\textbf{Implications of our results:} With our proposed theory, we can for the first time show that a maximum discovery latency can be guaranteed for all parametrizations except for a finite number of singularities, at which the latency converges towards infinity. This insight allows interval-based protocols even with $T_a > d_s$ to be applied in scenarios where deterministic maximum latencies are required, in which currently only slotted protocols are used.
In addition, our proposed theory can be used to find optimal parametrizations of protocols like ANT/ANT+ or BLE. Recent work based on this paper\footnote{This paper has been available as a preprint on arXiv.org since sept. 2015.} has shown that optimized PI-based protocols can significantly outperform all previously known discovery protocols, e.g., by achieving 10x shorter discovery-latencies than \textit{Searchlight} \cite{kindt:16} \cite{kindt:17a} \cite{kindt:17b}.

\noindent\textbf{Organization of the paper:} The rest of this paper is organized as follows. In Section \ref{sec:related_work}, we present related work. We formally define the problem of neighbor discovery in Section \ref{sec:problem_formulation}. In Section \ref{sec:overview}, we present our proposed theory for solving the neighbor discovery problem. In Section \ref{sec:algo}, we present an efficient implementation of our proposed technique. Readers who are only interested in understanding our proposed theory may only focus on Section \ref{sec:overview}. Only those wo are interested in an implementation may continue to Section \ref{sec:algo}. Next, we evaluate our theory with comprehensive discrete event simulations and real-world measurements in Section \ref{sec:evaluation}. In Section \ref{sec:parameter_selection}, we describe how the parameter values of interval-based protocols can be chosen in an optimal fashion. Finally, we summarize our results and discuss their implications in Section \ref{sec:conclusion}. 
%Based on the parameter valuations, our theory distinguishes between multiple cases. In the appendix, the solutions are presented for the case which have been left away due to space constraints. 
To ease the readability of this paper, a complete table of symbols used is given in Appendix~\ref{sec:table_of_symbols}.

\section{Related Work}
\label{sec:related_work}
In this section, we provide an overview on existing models for periodic interval-based, slotless protocols, such as BLE. As already mentioned, exact models only exist for the trivial case of $T_a < d_s$. This implies that the duration of one scan window must be larger than the time between two advertising packets. Therefore, every scan attempt is successful and the discovery latency is limited to roughly $1 \cdot T_s$. The first known model (for $T_a < d_s$) has been presented along with the introduction thew STEM-B protocol \cite{schurgers:02}, which is one of the first PI-based protocols that have been proposed. In \cite{liu:12_short,liu:12_long} and \cite{liu:12_techrep}, this solution has been adopted to the BLE protocol to account for multiple channels, but the limitation to $T_a < d_s$ remains. 

For the general case, as \cite{AntChannelSearch:09} states, there is no known model to compute the discovery latency of PI-based protocols. Simulations are computationally very complex, since the discovery process needs to be simulated repeatedly for a large number of initial time offsets in order to assess the mean discovery latencies. An attempt to reduce this complexity has been proposed in \cite{kindt:13}. However, the resulting complexity is still impractically high, and only an estimation of the mean latency can be given. No estimate on the maximum discovery latency can be provided.
An attempt to derive a probabilistic model for $T_a \geq d_s$ has been made in \cite{cho:15}, \cite{cho:15_2}. Here, the probability that an advertising packet meets a scan window has been assumed to be constant for all packets on a particular channel. In particular, if $p_m$ is the probability of a miss of the first advertising packet on a particular channel, then it is assumed that the miss-probability of the $k$'th advertising packet is $p_m^k$. However, the periodic nature of BLE implies that a strong correlation between the probability and the interval instance exists and hence $p_m(k)$ is a function of the interval count $k$. Hence, as we will also show in Section~\ref{sec:model_comparsion}, the results obtained from such models do not correlate well with the actual discovery latencies. Also, under the assumption of constant matching probabilities for all packets, a deterministic discovery is not guaranteed and therefore no upper bound can be inferred.
In \cite{matsuo:15}, the discovery latency of BLE has been computed by applying the probabilistic model checker PRISM~\cite{Kwiatkowska:11} for some discrete values of $T_a > d_s$, $T_s$ and $d_s$. However, in our opinion, the necessary details to understand this work have not been provided in \cite{matsuo:15}. In addition, the initial offset between the first advertising packet and scan window has been assumed to be fixed. As a result, this approach is unable to determine the mean- and worst-case discovery latencies among all possible initial offsets. 

Usually, PI-based protocols are considered to be non-deterministic, since no upper bound could be determined until now. 
To overcome the lack of determinism, it has been proposed to configure the BLE protocol such that the parameter values fulfill the CRT \cite{kandhalu:13}. However, such configurations restrict the range of possible valuations unnecessarily. In summary, to the best of our knowledge, no valid model for PI-based protocols for the general case ($T_a \geq d_s$) has been presented until now.

\noindent \textbf{Our contributions:} In this paper, we solve this problem and present the first mathematical theory of PI-based protocols. 
In particular, we make the following contributions:
\begin{asparaenum}
\item  We for the first time propose a model which can compute the exact mean and maximum discovery latencies for the complete range of possible parametrizations, including $T_a > d_s$. This theory leads to important new insights into such protocols and provides a full understanding of the probabilistic processes involved.
\item By using our model, we demonstrate that the discovery-latency is bounded for almost all parametrizations. The popular belief until now was that parametrizations with $T_a > d_s$ lead to unbounded discovery latencies. This new finding has important implications on the design of wireless protocols and the parametrization of protocols like BLE.
\item We evaluate our proposed solution both with comprehensive discrete-event simulations and real-world measurements, which are based on an implementation on two wireless radios. Thereby, we demonstrate that our predicted latency bound is safe (i.e., the latency never exceeds this bound) and tight (i.e., the bound can be reached in practice), and that our estimated mean latencies are precise.
\end{asparaenum}

 \section{Problem Formulation}
 \label{sec:problem_formulation}
 In this section, we formally define the problem that is addressed in this paper.
 \begin{figure}[htb]
 \centering
 \includegraphics[width = 1.0\columnwidth]{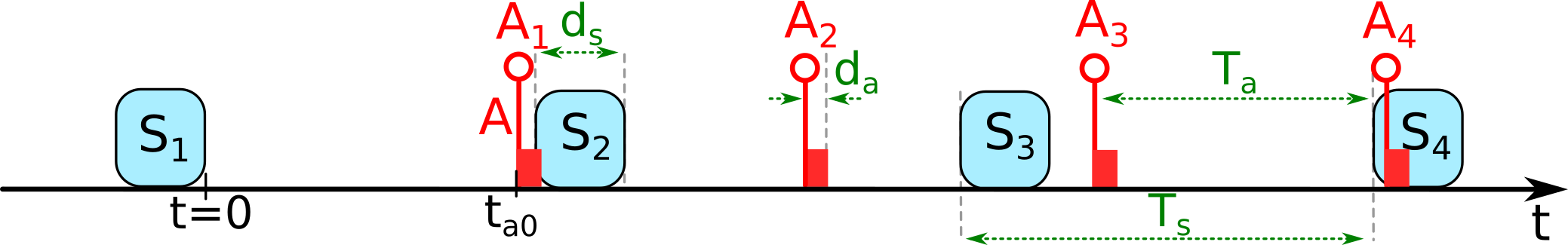}
 \caption{Sequence of advertising packets ($A_1-A_4$) and scan windows (${S_1-S_4}$) for neighbor discovery.}
 \label{fig:Problem Formulation} 
 \end{figure}
 One device, referred to as the \textit{advertiser} (A), periodically sends out \textit{advertising packets}, starting from an uniformly distributed\footnote{The assumption of uniformly distributed initial time-offsets influences the computed mean discovery latencies, but not the worst-case latencies.} random point in time $t_{a0}$. Transmitting a packet takes $d_a$ time-units. The packets are sent periodically with the \textit{advertising interval} $T_a$. The remote device is called the \textit{scanner} (S). It periodically switches on its receiver once per \textit{scan interval} $T_s$ and listens for advertising packets for a duration called the \textit{scan window} $d_s$. The neighbor discovery process is successful once an advertising packet has been received successfully by the scanner. Therefore, both devices need to be awake simultaneously for at least one packet transmission duration $d_a$. In Figure \ref{fig:Problem Formulation}, the first advertising packet (labeled with ($A_1$)) is not received successfully, as it only partially overlaps with the scan window $S_2$. The second ($A_2$) and third ($A_3$) advertising packets are not received either, because they do not overlap with any scan window at all. The fourth packet ($A_4$) lies entirely within the scan window $S_4$ and is therefore received successfully. We assume no packet loss and only one channel. Then, the necessary and sufficient condition for a successful reception is given by the existence of some $i,j \in \mathbb{N}_0$, such that
 \begin{equation}
 \label{eq:hit_condition}
 j \cdot T_s  - d_s \leq t_{a0} + i \cdot T_a \leq j \cdot T_s - d_a,
 \end{equation}
where $t_{a0}$ is the random point in time at which the first advertising packet is sent and $i$ and $j$ are the indices or the counting numbers of all advertising packets/scan windows.
The questions answered in this paper are the following.
What are the maximum and average times for an advertising packet to lie within a scan window for the first time (as given by Inequality \ref{eq:hit_condition}), i.e., what are the max and mean discovery latencies?
\vspace*{-1em}
\section{Protocol Analysis}
\label{sec:overview}
In this section, we describe the key concepts of our proposed solution to compute the mean- and maximum neighbor discovery latencies. To develop an intuitive understanding of this, we first present four different examples. Next, we formalize and generalize these examples. 
Towards this, without any loss of generality, we first set the duration of the advertising packets $d_a$ to $0$. Because a successful reception occurs only if the advertising packet is received entirely by the scan window, this measure can be compensated for by shortening the scan window $d_s$ by $d_a$ time units. Further, we have to add $d_a$ time units to all computed latencies.

\subsection{Examples}
\begin{figure}[tb]
	\centering
	\includegraphics[width=\linewidth]{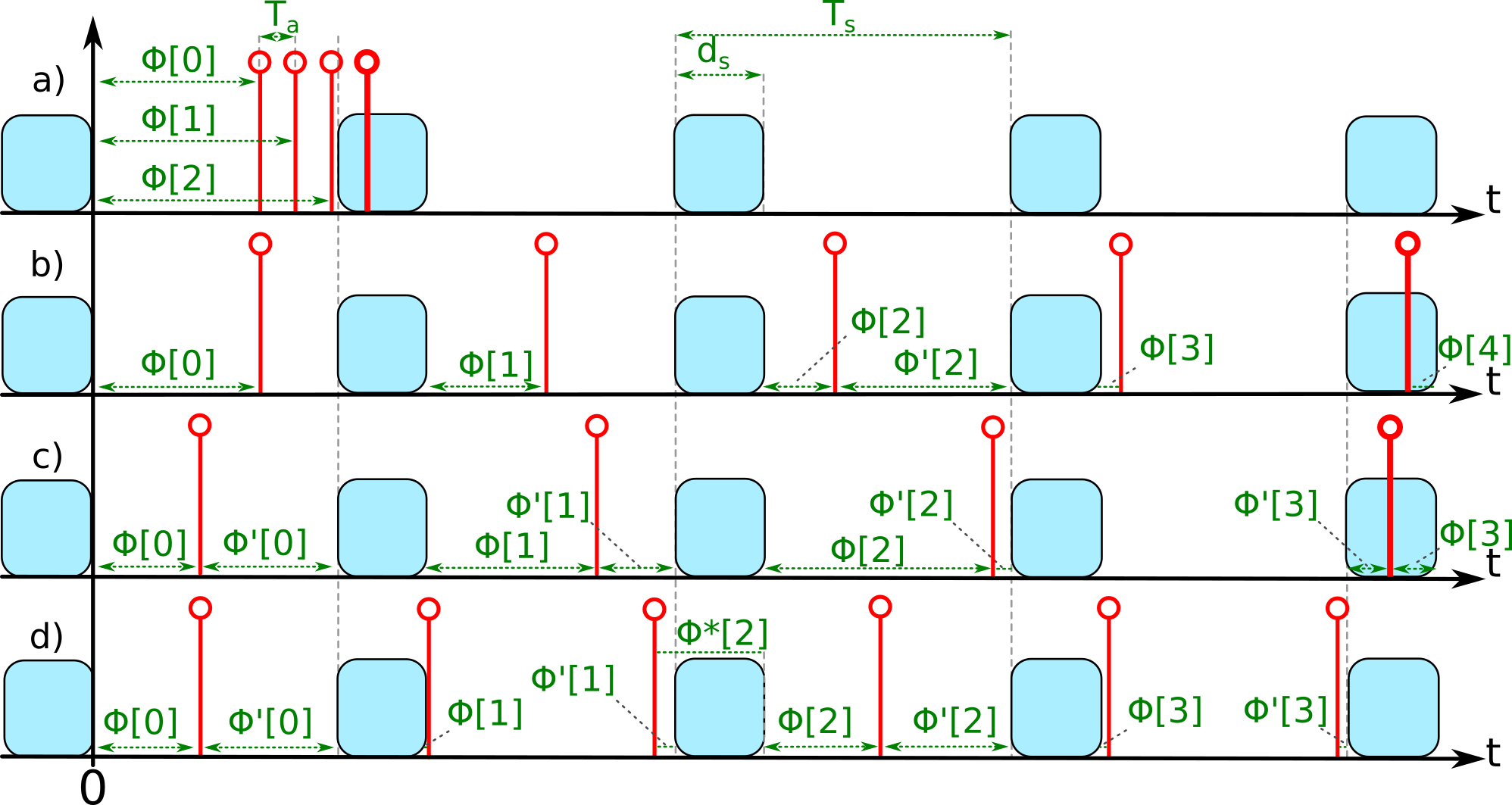}
	\caption{Examples for different situations of neighbor discovery processes. The rounded boxes depict the scan windows, whereas the lines with circles at their heads depict the advertising packets.}
	\label{fig:shrinkingGrowingConcept} 
\end{figure}

\noindent \textbf{a) $\mathbf{T_a \leq d_s}$:} 
Situations with $T_a \leq d_s$ form the most simple cases. Consider the situation depicted in Figure~\ref{fig:shrinkingGrowingConcept}~a). Again, the straight lines with circles on their heads depict the starting times of the advertising packets, whereas the boxes depict the scan windows. 
Here, $T_a \leq d_s$ and hence a successful match is guaranteed within one scan interval. The temporal distance between the second scan window and each advertising packet shrinks by $T_a$ time units with every packet sent. Given an initial offset $\Phi[0]$ between the end of the first scan window and the first advertising packet, the discovery-latency is defined by the number of advertising intervals that need to pass until the scan window is reached. It follows that the neighbor discovery latency $d_{nd}$ is:
 \begin{equation}
 \label{eq:solutionFigConceptA}
d_{nd}(\Phi[0])   = \left\{\begin{array}{ccl}  \left\lceil \frac{T_s - \Phi[0] - d_s}{T_a}\right\rceil \cdot T_a, & \mbox{if} & \Phi[0] \leq T_s - d_s,\\
	0, & & \mbox{otherwise.}  \\
	 \end{array}\right.
 \end{equation}

As can be seen easily from the figure, the maximum discovery latency for this example is $d_{nd,m} = \lceil \frac{T_s - d_s}{T_a} \rceil \cdot T_a$.

\noindent \textbf{b) $\mathbf{T_a > d_s}$:}
Whenever $T_a > d_s$, as depicted in Figure~\ref{fig:shrinkingGrowingConcept}~b), a successful discovery within one scan interval is not guaranteed anymore. Here, it is beneficial to describe the problem using a different representation.
Instead of considering the absolute points in time the advertising packets and scan windows begin at, we examine the relative time differences $\Phi[k]$ between each scan window $k=0,1,2,...$ and its neighboring advertising packets. As we will introduce later, there are ``shrinking'' and ``growing'' constellations of scan windows and advertising packets. The example in Figure~\ref{fig:shrinkingGrowingConcept}~b) is shrinking, and here only each closest neighboring advertising packet that is temporally on the right of a scan window (i.e., the first advertising packet on the right of each scan window) is relevant. 

From Figure~\ref{fig:shrinkingGrowingConcept}~b), one can observe that in this example, the temporal distance $\Phi[k]$ is monotonically shrinking for increasing indices $k$, and the amount of shrinkage is constant per scan interval $k$. This shrinkage is denoted as $\gamma_0$, and it is $\gamma_0 = \Phi[k] - \Phi[k + 1]$. Here, one can observe that a match occurs after $\Phi[k]$ becomes smaller than $0$ (cf. $\Phi[4] < 0$ in Figure~\ref{fig:shrinkingGrowingConcept}~b). This holds true for any initial offset, as long as the shrinkage $\gamma_0$ is smaller or equal than $d_s$ .

In the previous example with $T_a \leq d_s$, (cf. Figure \ref{fig:shrinkingGrowingConcept}~a)), we had counted the number of advertising intervals that fit into the temporal distance to the next scan-window (i.e., ${T_s - d_s - \Phi[0]}$ time units) for computing the discovery latency. The example from Figure \ref{fig:shrinkingGrowingConcept}~b) can be handled in the same way - with the only difference being that we have to count the number of $\gamma_0$-intervals that fit into the initial offset, instead of the number of advertising intervals. 
Hence, the discovery latency is defined by Equation~\eqref{eq:ni}.

 \begin{equation}
 \label{eq:ni}
 d_{nd}(\Phi[0])  = \left\{\begin{array}{ccl} \left\lceil \frac{\Phi[0] - d_s}{\gamma_0} \right\rceil \cdot T_a& \mbox{, if} & \Phi[0] \geq d_s \\
	0 & \mbox{, else.} &  \\
	 \end{array}\right.	
 \end{equation}

\noindent\textbf{c) $\mathbf{T_a > d_s}$, growing:}
In the previously examined situation (depicted in Figure~\ref{fig:shrinkingGrowingConcept}~b)), the offset $\Phi[k]$ shrank for increasing values of $k$. However, there are cases in which $\Phi[k]$ becomes larger with increasing values of $k$, as depicted in Figure~\ref{fig:shrinkingGrowingConcept}~c). Whereas the distance from a scan window $k$ to its right neighboring advertising packet $\Phi[k]$ grows with increasing values of $k$, the distance to its left neighboring advertising packet $\Phi'[k]$ shrinks. Growing constellations can therefore be handled similarly to shrinking ones.

\noindent\textbf{d) $\mathbf{\gamma_0 > d_s}$:}
The examples b) and c) both had $\gamma_0 \leq d_s$ in common. However, for $\gamma_0 > d_s$, the computation becomes more involved. In the example from Figure~\ref{fig:shrinkingGrowingConcept}~d), the distance $\Phi[0]$ shrinks from $\Phi[0]$ to $\Phi[1]$ by $\gamma_0$ time units. After $\Phi[1]$, this distance $\Phi^*_2$ (we label this distance with *, because it is no longer the distance to the closest advertising packet on the right of the scan window) becomes smaller than $-d_s$, since the advertising packet has moved to the left side of the scan window. Because advertising packets on the left have no chance of hitting the scan window considered (given the situation is shrinking), the advertising packet that needs to be considered changes (e.g., from $\Phi[2]^*$ to $\Phi[2]$ in the figure). One advertising packet is skipped and the offset $\Phi[k]$ becomes larger again for this packet (i.e. $\Phi[2] > \Phi[1]$).

We say that after $\Phi[1]$, the advertising packet ``overtakes'' the scan window. After that, from $\Phi[2]$ to $\Phi[3]$, the distance shrinks again by $\gamma_0$, until the relevant packet changes the next time. Shrinking the initial offset successively by $\gamma_0$ time units cannot guarantee discovery, since the ``overtaking'' might repeat. However, let us consider two consecutive skipped advertising packets, e.g., the packets with the offsets $\Phi'[1]$ and $\Phi'[3]$ in Figure~\ref{fig:shrinkingGrowingConcept}~d). Both packets fulfill the condition that they are the first ones after an advertising packet has ``overtaken'' the corresponding scan window, and it is $\Phi'[1] - \Phi'[3] = const. = \gamma_1 < \gamma_0$.
In other words, when considering appropriate multiples of the advertising interval (here: $i_2 = 3 \cdot T_a$) and scan interval (here: $j_2 = 2 \cdot T_s$), given an offset $\Phi[k]$ between a scan window $k$ and its neighboring advertising packet, the offset between a scan window that is $i_n$ scan intervals and $j_n$ advertising intervals later will be $\Phi[k] - \gamma_1$. In case \ref{fig:shrinkingGrowingConcept} d), $\gamma_1 < d_s$ and therefore the advertising packets and scan windows will match guaranteed after the initial offset $\Phi[0]$ has shrunk by a certain number of steps of length $\gamma_1$. 

\subsection{$\gamma$-Sequences}
\label{sec:gammaSequences}
We now generalize the above examples. Towards this, let us first define the concept of $\gamma$-sequences and state multiple properties, which we claim to be true. We will then use these properties to compute the neighbor discovery latencies for arbitrary parametrizations. We will focus our attention on proving these claims in Appendix \ref{seq:proofs}.

\begin{definition}{\textbf{$\mathbf{\gamma}$-sequences:}} 
	\label{def:gammaSequence}
Let there be a set of parameters $(T_a, T_s, d_s)$ and pairs of integers ($i_n$, $j_n$), $n = 0,1,2,...n_m$, with $i_{n} \geq i_{n-1}$ and $j_n \geq j_{n-1}$. We call a sequence of scan windows and advertising packets defined by ($1 \cdot i_n  \cdot T_a,\mbox{ } 1 \cdot  j_n \cdot T_s),\mbox{  } (2 \cdot  i_n  \cdot T_a,\mbox{ } 2 \cdot j_n  \cdot T_s),...$ a $\mathit{\gamma}$-\textit{sequence} of order $n$, if the following properties are fulfilled.

\begin{property} 
\label{prop:constantShrinkage}
If the time offset between any scan window and advertising packet is $\Phi$, the time offset between the scan window that is $j_n$ scan-intervals later and the advertising packet that is $i_n$ advertising intervals later will be $\Phi \pm \gamma_n$. Sequences of tuples of scan windows and advertising packets, in which such offsets always decrease, are called  \textit{shrinking} $\gamma$-sequences, whereas constellations, in which these offsets increase, are called \textit{growing} $\gamma$-sequences.
\end{property}

\begin{property} 
\label{prop:rangeOfGamma}
$0 \leq \gamma_n \leq \min(T_a, T_s),\mbox{ }\forall n$
\end{property}

\begin{subdefinition}{\textbf{Mode $\mathbf{m}$}:} 
	\label{def:mode}
	We define $m_n \in (g,s,c)$ as the \textit{mode} of a $\gamma$-sequence of order $n$. If the offset $\Phi$ between any scan window and any advertising packet becomes larger every $i_n$ advertising packets and $j_n$ scan-windows, then $m_n = g$ (growing). If it becomes smaller, then $m_n = s$ (shrinking). If $\gamma_n = 0$, then there is no shrinkage or growth, and $m_n = c$ (coupling).
\end{subdefinition}

\begin{property}
\label{prop:gammaRecursiveDefinition}
$\gamma_{n} = | T_s - |\sum_{k = 0}^{n-1} a_k \cdot \gamma_{k}||$, $a_k \in \mathbb{Z} \setminus 0$. Here, the only valid choices of $a_k$ are those that satisfy the following: 
i) $\gamma_{n} \leq \frac{1}{2} \gamma_{n-1}$, ii) the sum $\sum_{k=0}^{n-1} |a_k| \cdot (i_k T_a)$ is minimized, and iii) for a shrinking sequence, $a_k < 0$, otherwise, $a_k > 0$.
\end{property}

\begin{property} 
	\label{prop:lowestOrderFirst}
For any $\gamma_n$, $i_n \cdot T_a$ has always at least the value of the smallest possible sum $\sum_{k=0}^{n-1} |a_k| \cdot (i_k T_a)$, for which $|\sum_{k=0}^{n-1} a_k \gamma_k| \geq T_s - \gamma_n \geq T_s - \frac{1}{2} \gamma_{n-1}$.
\end{property}

\begin{property} 
	\label{prop:limitedOrder}
     The smallest $n$ for which $\gamma_n \leq d_s$ is called the \textit{maximum order} $n_m$ for the given parametrization. It is
	\begin{equation}
	\label{eq:maxIterationOrder}
	n_m = \left \{ \begin{array}{lll}\left\lceil \frac{\ln(\min(T_a,T_s))-\ln(d_s)}{\ln(2)} \right\rceil, & \mbox{if} & \min(T_a,T_s) > d_s, \\ 0,& &\mbox{otherwise.} \end{array}  \right.
	\end{equation}
\end{property}

\begin{subdefinition}{\textbf{Penalty $\mathbf{\sigma}$}:} 
	\label{def:penalty}
 Let there be a $\gamma$-sequence with shrinkage/growth $\gamma_n$. Then $\sigma_n = i_n \cdot T_a$ is called the \textit{penalty} for increasing or decreasing any offset between a scan window and an advertising packet by $\gamma_n$ time units.
 \end{subdefinition}
 \begin{lemma}
 	\label{lemma:penaltyValue}
 	Property~\ref{prop:gammaRecursiveDefinition} implies that $\sigma_n =\sum_{k = 0}^{n-1} |a_k| \cdot \sigma_{k} = i_n \cdot T_a$, where $a_k$ are equal to the coefficients in Property~\ref{prop:gammaRecursiveDefinition}. Further, $\sigma_{n} > \sigma_{n-1}$, since $i_n > i_{n-1}$.
 \end{lemma}
\end{definition}

Let us assume that the properties above hold true. We can then make use of them to solve the neighbor discovery problem, as defined by Inequality \eqref{eq:hit_condition}, as follows.

\begin{asparaenum}
	\item First, $\gamma_0,\gamma_1,...,\gamma_{n_m}$ need to be computed. This computation is described in detail in Section \ref{sec:computingGamma}.
	\item Next, the entire range of possible initial offsets is subdivided into partitions with equal latencies. For each partition, the latency can be derived by identifying a set of $\gamma$-sequences that leads to a coincidence of a packet and a corresponding scan window. This procedure is described in Section \ref{sec:partitioning}.
	\item From the set of partitions with corresponding latencies, one can extract the global worst-case and mean latencies.
\end{asparaenum}

\subsection{Computation of the $\gamma$ - Parameter}
\label{sec:computingGamma}
This section is dedicated to the computation of $\gamma_n$.
Property~\ref{prop:gammaRecursiveDefinition} implies that $\gamma_n$ is defined by the absolute difference between $T_s$ and the absolute value of a linear combination of lower-order $\gamma$-values, such that $i_n$ (and hence $\sigma_n$) is minimized, under the constraint that $\gamma_n < \frac{1}{2} \gamma_{n-1}$. We in the following present a recursive scheme to compute $\gamma_n$ given $\gamma_{k},\mbox{ }k \in 0..n-1$. First, we present the computation of the initial value $\gamma_0$ of this recursion.
\subsubsection{Initial Values}
For realizing a shrinkage or growth of $\gamma_n$ time units every $i_n$ advertising packets and $j_n$ scan windows, the amount of time defined by $i_n \cdot T_a$ must be by $\gamma_n$ time units shorter (for a shrinkage) or longer (for a growth) than $j_n \cdot T_s$. Hence, we have to identify the smallest $i_n$ and a suitable $j_n$, such that $|i_n T_a - j_n T_s| = \gamma_n < \min(T_a, T_s)$ (cf. Property~\ref{prop:rangeOfGamma}). 
If $T_a < T_s$, every single advertising packet is part of a growing $\gamma$-sequence with $i_0 = 1, j_0 = 0$, and it is $\gamma_0 = T_a$.
If $T_a > T_s$, the $\gamma$-sequence of order 0 implies $i_n = 1$, since this is the lowest possible multiple of $T_a$. There are two possible values for $j_n$, such that Property \ref{prop:rangeOfGamma} (i.e., $\gamma_0 < T_s$) is fulfilled. The first one is defined by the smallest multiple of $T_s$ that exceeds $T_a$, which forms a shrinking sequence with $\gamma_{s,0}$. The other one is defined by the largest multiple of $T_s$ that does not exceed $T_a$, which forms a growing Sequence with $\gamma_{g,0}$. It is
\begin{equation}
\label{eq:gammaDefSh}
\gamma_{s,0} = \left\lceil\frac{T_a}{T_s}\right\rceil \cdot T_s - T_a, \mbox{ }
\gamma_{g,0} = T_a - \left\lfloor\frac{T_a}{T_s}\right\rfloor \cdot T_s.	
\end{equation}
Though both $\gamma_{s,0}$ and $\gamma_{g,0}$ are valid values for $\gamma_0$, we choose $\gamma_0 = \min(\gamma_{g,0}, \gamma_{s,0})$, and set $m_0 = s$ or $g$, accordingly, because this will result in a lower highest order $n_m$.
Next, we describe how to compute $\gamma_n$ with $n > 0$ .
 
\subsubsection{Recursion Scheme}

\begin{figure}[b]
	\centering
	\includegraphics[width = 1.0\columnwidth]{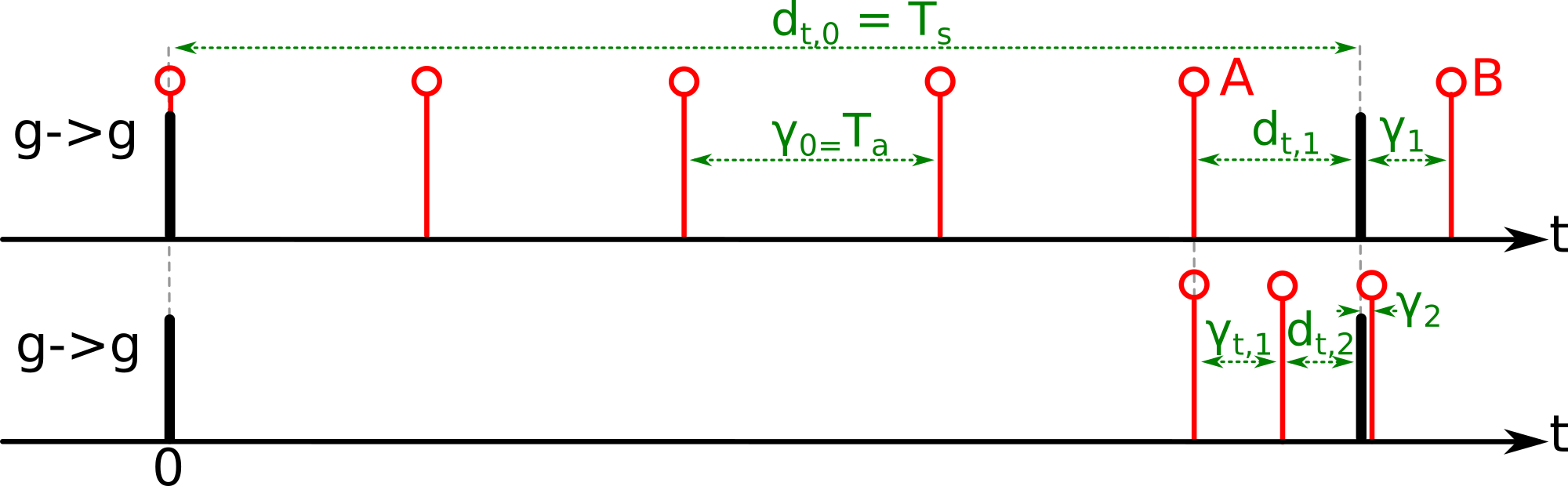}
	\caption{Computation of $\gamma_2$ given $\gamma_1$ for growing sequences.}
	\label{fig:computegamma_grgr} 
\end{figure}

We in the following assume two sequences with $m_{n-1} = m_{n} = g$, as depicted in Figure~\ref{fig:computegamma_grgr}, and restrict our description to this. We generalize it afterwards.

Let $\gamma_{n-1}$ describe a certain offset-shrinkage or -growth (cf. Property \ref{prop:constantShrinkage}). Since this shrinkage or growth applies for the offset between any pair of scan windows and advertising packets, also offsets that result from a previous shrinkage/growth are again shrunk or grown by $\gamma_{n-1}$ time units, whenever the additional advertising and scan intervals pass. Hence, linear combinations ${L_{\gamma_n} = \sum_{k=0}^{n} a_k\gamma_{k}}$, with $a_k \geq 0$, if $m_{k} = g$ and $a_k < 0$, if $m_{k} = s$, will result in an effective growth or shrinkage of $L_{\gamma_n}$ time units.
Further, if $L_{\gamma_n} > 0$, we can form differences $L_{\gamma_n} - T_s$ and obtain an effective shrinkage or growth $\gamma_n' = |L_{\gamma_n} - T_s|$. If $L_{\gamma_n}$ involves $i_n$ advertising intervals and $j_n$ scan intervals, $\gamma_n'$ can be regarded as the shrinkage (if $L_{\gamma_n} - T_s < 0$) or growth (if $L_{\gamma_n} - T_s > 0$) after every $i_n$ advertising intervals and $j_n + 1$ scan intervals. Similarly, if $L_{\gamma_n} < 0$,  $\gamma_n' = L_{\gamma_n} + T_s$ is the shrinkage (if  $L_{\gamma_n} < - T_s$) or growth (if $L_{\gamma_n} > - T_s$) after $i_n$ advertising and $j_n+1$ scan intervals.

We can use this to construct $\gamma_{n} = | T_s - |\sum_{k = 0}^{n-1} a_k \cdot \gamma_{k}||$, $a_k \in \mathbb{Z}  \setminus 0$, by choosing $a_k$ such that $|L_{\gamma_n}| = |\sum_{k = 0}^{n-1} a_k \cdot \gamma_{k}|$ differs by less than $\frac{1}{2} \cdot \gamma_{n-1}$ time-units from $T_s$ and hence, ${\gamma_{n} < \frac{1}{2} \gamma_{n-1}}$. There are different choices of $a_k$ that fulfill $\gamma_n < \frac{1}{2} \cdot \gamma_{n-1}$, and we have to identify the one that leads to the lowest penalty $\sigma_n = \sum_{k = 0}^{n-1} |a_k| \cdot \sigma_{k}$. Property~\ref{prop:lowestOrderFirst} implies that shrinking or growing any offset by $\gamma_n$ time-units always incurs at least the penalty $\sigma_n$ of the linear combination $|\sum_{k=0}^{n-1} a_k \gamma_k|$ that i) reaches or exceeds $T_s - \frac{1}{2}\gamma_{n}$, and ii) has the lowest possible corresponding sum $\sum_{k=0}^{n-1} |a_k| \sigma_k$. Since we attempt to find the ``best'' (in terms of penalty) linear combination $|\sum_{k=0}^{n-1} a_k \gamma_k| > T_s - \frac{1}{2}\gamma_{n-1}$, Property~\ref{prop:lowestOrderFirst} implies that one instance of $\gamma_{n-1}$ in such a linear combination incurs already the penalty of the ``best'' linear combination that contains no instance of $\gamma_{n-1}$, if such a combination exists. As a result, coefficients $a_k$ for higher sequence orders $k$ always need to be minimized at the cost of larger coefficients $a_k$ for lower orders $k$.

In the example depicted in Figure~\ref{fig:computegamma_grgr}, $\gamma_1$ and $\gamma_2$ are computed given $\gamma_0$. Here, $m_1 = m_2 = g$, and therefore ${L_{\gamma_{n-1}}}$ must always exceed $T_s$. Hence, it is $\gamma_{n} = L_{\gamma_{n-1}} - T_s$. For computing $\gamma_1$, let us find the lowest multiple $Q_0$ of $\gamma_0$ time units, such that this multiple exceeds $T_s$. We can regard this as finding the number of $\gamma_0$ time-intervals that ``fit'' into a certain \textit{distance to travel} $d_{t,0} = T_s$, i.e., ${Q_0 = \lfloor \frac{d_{t,0}}{\gamma_0} \rfloor}$. As can be seen in Figure \ref{fig:computegamma_grgr}, 
$\gamma_{1} = (Q_0+1) \gamma_{0} - T_s$, since this is the linear combination with the lowest sum of penalties (here: $(Q_0+1) \cdot \sigma_0$) that fulfills $\gamma_1 \leq \frac{1}{2} \gamma_0$.

For computing $\gamma_2$, we consider the largest multiple $Q_0$ of $\gamma_0$-intervals that does not exceed $T_s$. The remaining difference from $T_s$ is $d_{t,1} = d_{t,0} - Q_0 \gamma_0$. We now identify the largest multiple of $\gamma_1$ time units that  ``fit'' into $d_{t,1}$, i.e., ${Q_1 = \lfloor \frac{d_{t,1}}{\gamma_1} \rfloor}$. Using this, ${\gamma_2 = Q_0 \gamma_0 + (Q_1+1) \gamma_1 - T_s}$.
Any linear combination that involves a higher number of $\gamma_1$ time units than $Q_1$ would violate Property \ref{prop:lowestOrderFirst}. This scheme can be generalized for arbitrary orders $n$. With $Q_n = \lfloor \frac{d_{t,n}}{\gamma_n} \rfloor$,  it is
\begin{equation}
\label{eq:gammaNextGrGr}
\begin{array}{lcl}
\gamma_n		&	=	&	(Q_{n-1} + 1) \cdot \gamma_{n-1} - d_{t,n-1},\\
d_{t,n}			&	=	&	d_{t,n-1} - Q_{n-1} \cdot \gamma_{n-1},\\
\sigma_n		&	=	&	\sigma_{s,n-1} + (Q_{n-1}+1) \cdot \sigma_{n-1},\\
\sigma_{s,n}	&	=	&	\sigma_{s,n-1} + Q_{n-1} \cdot \sigma_{n-1}.
\end{array}
\end{equation}
The first distance to travel $d_{t,0}$ is initialized by $T_s$, $\sigma_{s,0}$ is initialized by $0$ and $\sigma_0$ by $T_a$. Here, $\sigma_{s,n}$ is the penalty (i.e., sum of advertising intervals) for any linear combination $\sum_{k=0}^{n-1} a_k \gamma_k$ that is by $d_{t,n}$ smaller than $T_s$. 

In fact, one can show that Equation \ref{eq:gammaNextGrGr} holds true whenever $m_{n} = m_{n-1}$ (i.e., also for $m_{n} = s = m_{n-1}$). Similar considerations also exist for $m_{n} \ne m_{n-1}$, which lead to slightly different equations. The interested reader may refer to Appendix \ref{app:recursiveGammaComputation}. Depending on $m_{n}$ and $m_{n-1}$, one out of two sets of equations has to be applied. This requires the mode $m_{n}$, which can be computed given $\gamma_{n-1}$, $m_{n-1}$ and $d_{t,n_1}$, as defined in Table \ref{tab:m_N}.

\begin{table}
	\begin{center}
		\caption{Recursive definition of the sequence mode $m_n$ given $m_{n-1}$}
		\label{tab:m_N}
		\renewcommand{\arraystretch}{0.9}% Tighter
		\begin{tabular}{cccc}
			$d_{t,n-1} - \lfloor \frac{d_{t,n-1}}{\gamma_{n-1}} \rfloor$ & $ < \sfrac{1}{2} \cdot \gamma_{n-1}$ &$ > \sfrac{1}{2} \cdot \gamma_{n-1}$ & $ = \sfrac{1}{2} \cdot \gamma_{n-1}$ \\ 
			\hline $m_{n-1} = s$ & g & s & c \\
			\hline $m_{n-1} = g$ & s & g & c \\
			\hline $m_{n-1} = c$ & - & - & - \\
		\end{tabular}
	\end{center}
\end{table}

\subsection{Latency Computation} 
 \label{sec:partitioning}

The first packet sent by a device that comes into range of a second, scanning device might fall into any part of its scan interval with an equal probability. Hence, the valid range of initial offsets between a scan window and an advertising packet is restricted to \mbox{$\Phi[0] \in [0, T_s]$}. For computing the discovery latency for a given $\Phi[0]$, we have to identify the linear combination $L' = \sum_{k=0}^{n_m} a_k \cdot \gamma_k$ that either fulfills ($- d_s \leq \Phi[0] + L' \leq 0$, or $T_s - d_s \leq \Phi[0] + L' \leq T_s$, and at the same time minimizes the corresponding sum of penalties ${\sum_{k=0}^{n_m} a_k \cdot \sigma_n}$. Minimizing the sum of penalties is required, because a linear combination with a higher sum of penalties would relate to a later coincidence of an advertising packet and a scan window, whereas the discovery latency is given by the first rendezvous. The intuition behind the latency computation procedure, along with an example, is given here. It is followed by the full algorithm in Section \ref{sec:algo}.

As already explained, a $\gamma$-sequence incurs a shrinkage or growth of the temporal offset $\phi$ between any scan window and any advertising packet after every $i_n$ advertising and $j_n$ scan intervals. Equivalently, one could describe this as \textit{one} advertising packet being ``shifted'' by multiples of $\gamma_n$ towards the left (if $m_n = s$) or the right (if $m_n = g$) within \textit{one} scan-interval. By summing up the penalties $\sigma_n$ of every ``shift'' of length $\gamma_n$, one can account for the fact that each shrinkage or growth actually occurs for later pairs of scan windows and advertising packets.
With this representation, the problem of finding such linear combinations $L$ can be solved by finding the ``best'' (in terms of penalties) linear combination of ``shifts'' of length $\gamma_n$, until the advertising packet reaches either the scan window on the left or on the right of it. Again, coefficients $a_k$ for high orders $k$ always have to be minimized at the cost of lower-order ones, as proven in Appendix~\ref{sec:latencyCompProof}.
\begin{asparaenum}
\item First, we create a set of partitions, such that the corresponding scan window (i.e., the left one for $m_n = s$ or the right one for $m_n = g$) is either reached or ``overtaken'' after the same multiple of $\gamma_0$ time units from within each partition. 
\item Each such partition is then subdivided in two other partitions. One of them contains all initial offsets from which the scan window can be reached (without ``overtaking'') after a multiple of $\gamma_n$ time units. We call them \textit{complete} partitions, since they do not need to be processed further, because discovery occurs after the corresponding multiple of $\gamma_0$ time units. The other ones contain the remaining offsets.
\item The partitions with the remaining offsets need to be subpartitioned again, such that after the same number of $\gamma_1$ time units, either the closest scan window or complete partition on the left (if $m_1 = s$) or right (if $m_1 = g$ is reached) or ``overtaken''.
\item Steps 2) and 3) are repeated for higher orders $n$, until $\gamma_n < d_s$. 
\end{asparaenum}

\begin{figure}[bth]
	\centering
	\includegraphics[width=\linewidth]{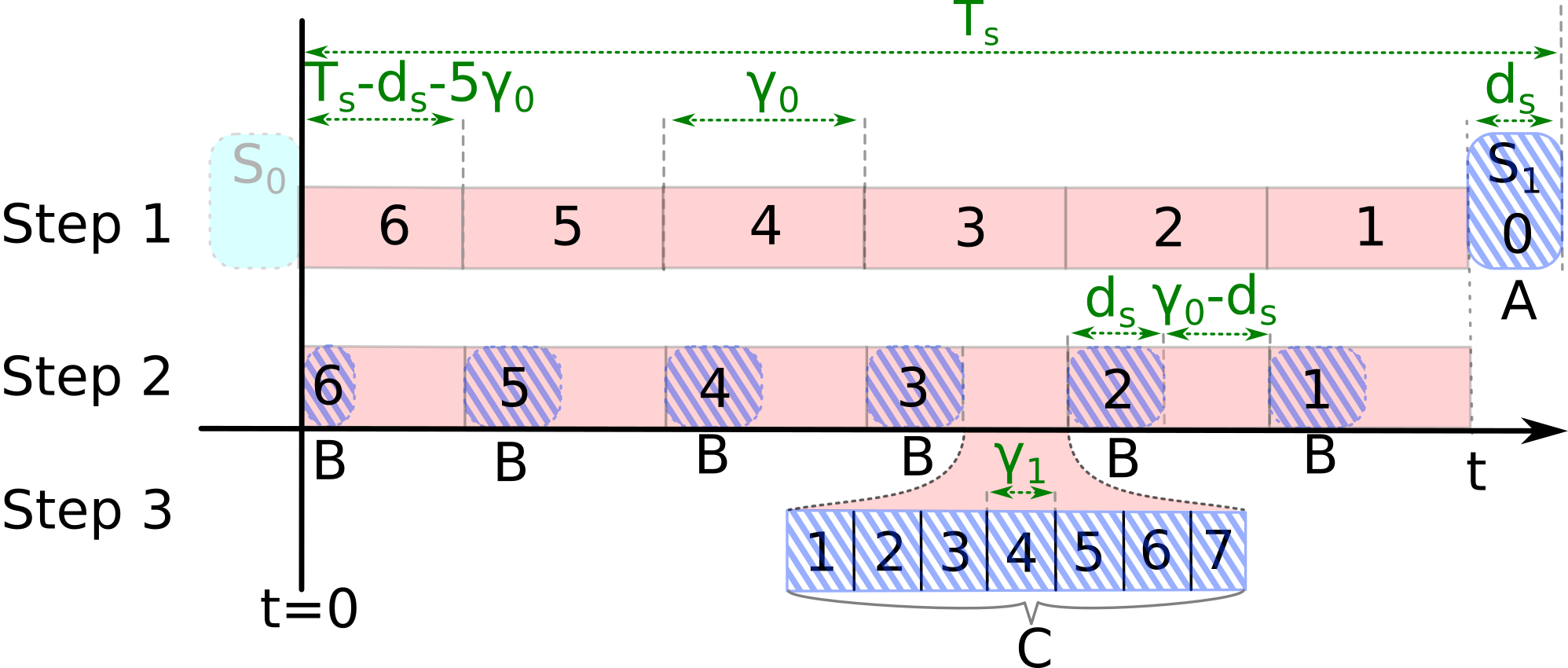}
	\caption{Computing the mean discovery latency for a higher order sequence.}
	\label{fig:modelHigherOrderScheme} 
\end{figure}

 In the following, we illustrate this scheme using an example. 
 Figure~\ref{fig:modelHigherOrderScheme} exemplifies a partitioning of the entire range of initial offsets for a growing sequence with parameter $\gamma_0$ and a shrinking sequence with $\gamma_1$. In Figure~\ref{fig:modelHigherOrderScheme}, the hatched rounded rectangles depict the complete partitions, whereas the remaining rectangles depict temporary partitioning steps that need to be further subdivided later. 
 In \textbf{Step 1} (cf. Figure~\ref{fig:modelHigherOrderScheme}), we subdivide the entire range $[0, T_s]$ into partitions of possible valuations of initial offsets, from which the scan window $S_1$ can be reached or ``overtaken'' using same multiple of $\gamma_0$ time units (i.e., the number of $\gamma_0$-intervals represented by the digits in each rectangle in Figure~\ref{fig:modelHigherOrderScheme}). A special case of this partitioning scheme is the scan window $S_1$ itself (cf. \textbf{A} in Figure~\ref{fig:modelHigherOrderScheme})), since an an advertising packet falling into this scan window has zero latency.  

In \textbf{Step 2}, each of these partitions is further subdivided into parts from which a multiple of $\gamma_1$ time units can reach the scan window $S_1$ (i.e., the hatched boxes in Figure \ref{fig:modelHigherOrderScheme}), and the remaining parts, from which $S_1$ would be ``overtaken''. For the hatched parts denoted with \textbf{B} in the figure, no further partitioning is needed, since the discovery latency for an initial offset falling into such a part is defined by the number of $\gamma_0$-intervals until reaching the scan window $S_1$. These partitions are called \textit{complete}.

However, the non-hatched parts from Step 2, which would ``overtake'' $S_1$, need to be further subdivided in \textbf{Step 3}.
Each of them is subdivided into multiple partitions of length $\gamma_1$ (except for the leftmost one, which is shortened to fit within the $\gamma_0 - d_s$ time units of the containing partition obtained from Step 2). These partitions are marked with \textbf{C} in the figure. In each of these partitions, the number of $\gamma_1$ - steps a packet has to be ``shifted'' by to reach its closest complete partition (or $S_0$  from the leftmost partition, respectively) is equal. Once a complete partition has been reached (or $S_0$ from the leftmost partition), $S_1$ is approached in steps of $\gamma_0$, as described before. Since $\gamma_1 < d_s$, no parts of these partitions can ``overtake'' the corresponding partition of type \textbf{B} (or the scan-window $S_0$, respectively). Therefore, the partitioning process is complete.

The discovery latency of any partition can be computed by summing up the appropriate penalties. For every partition, we can count the number of $\gamma_n$ - intervals until reaching the scan-window ($S_1$ in Figure \ref{fig:modelHigherOrderScheme}). For example, from  a partition of type \textbf{C}, we count the number of $\gamma_1$-intervals $n_1$ until reaching a partition of type \textbf{B}. From there, we count the number of $\gamma_0$-intervals $n_0$ until reaching the scan window $S_1$. The digits in each partition in Figure \ref{fig:modelHigherOrderScheme} correspond to $n_0$ and $n_1$. The discovery latency $d_{nd}$ is then $n_2 \cdot \sigma_2 + n_1 \cdot \sigma_1$.

This scheme, together with Property~\ref{prop:limitedOrder}, also implies that a scan window can always be reached within a finite number of steps. Therefore, the worst-case latency is always bounded, if $m_n \ne c \mbox{ }\forall n < n_m$.

\section{Implementation of the Model}
\label{sec:algo}

 A straight-forward but computationally complex approach for computing the mean and worst-case neighbor discovery latencies would be iterating over every single partition and computing the latencies for each of them, as outlined in the previous section.
 However, the computation can be done more efficiently, without carrying out the entire partitioning explicitly. In what follows, we present an efficient algorithm for computing the discovery latency, which is based on two main concepts:
1) Discrete summations: Instead of iterating over every single partition of a given sequence order, we account for multiple of them at once by appropriate discrete summations.
2) Separation of sequence orders: Rather than carrying out the entire partitioning for all $\gamma$-sequences, we first only perform the partitioning for a single sequence order. After each order, the results are merged, since the resulting latency distributions are identical for many of the processed partitions. Therefore, the computational complexity of the next iteration is reduced.
We provide a ready-to-use MATLAB-implementation of this algorithm for download (cf. Appendix~\ref{sec:using_the_model}).

The algorithm we propose handles probability distributions that consist of multiple probability values over piecewise constant time intervals. For this purpose, we define a data structure called a \textit{probability buffer} $\Xi$. It is described next.

\subsection{Probability Buffers}

A \textit{probability buffer} consists of a set of $k$ probability densities $\Xi[k]$, which are called \textit{segments}. Each of them is defined over a unique, disjoint time-interval, which consists of a start time $t_s[k]$ and an end time $t_e[k]$. The elements are sorted by their starting times $t_s[k]$ in an ascending order. 
We define $\| \Xi \|$ as the number of segments in the probability buffer $\Xi$.
For each probability buffer, we define an operation \textit{add($t_{ss}$, $t_{ee}$, $p$)}, which creates a new segment from $t_{ss}$ to $t_{ee}$ with probability $p$. Existing probability segments that lie at least partially within $[t_{ss}, t_{ee}]$ are split and their probabilities are added for the overlapping time duration. To add a segment from $t_{ss}$ to $t_{ee}$ with probability $p$, we introduce the following notation: $[t_{ss}, t_{ee}] \gets p$.

\subsection{Example}
\begin{figure}[tbh]
\centering
\includegraphics[width=1.0\columnwidth]{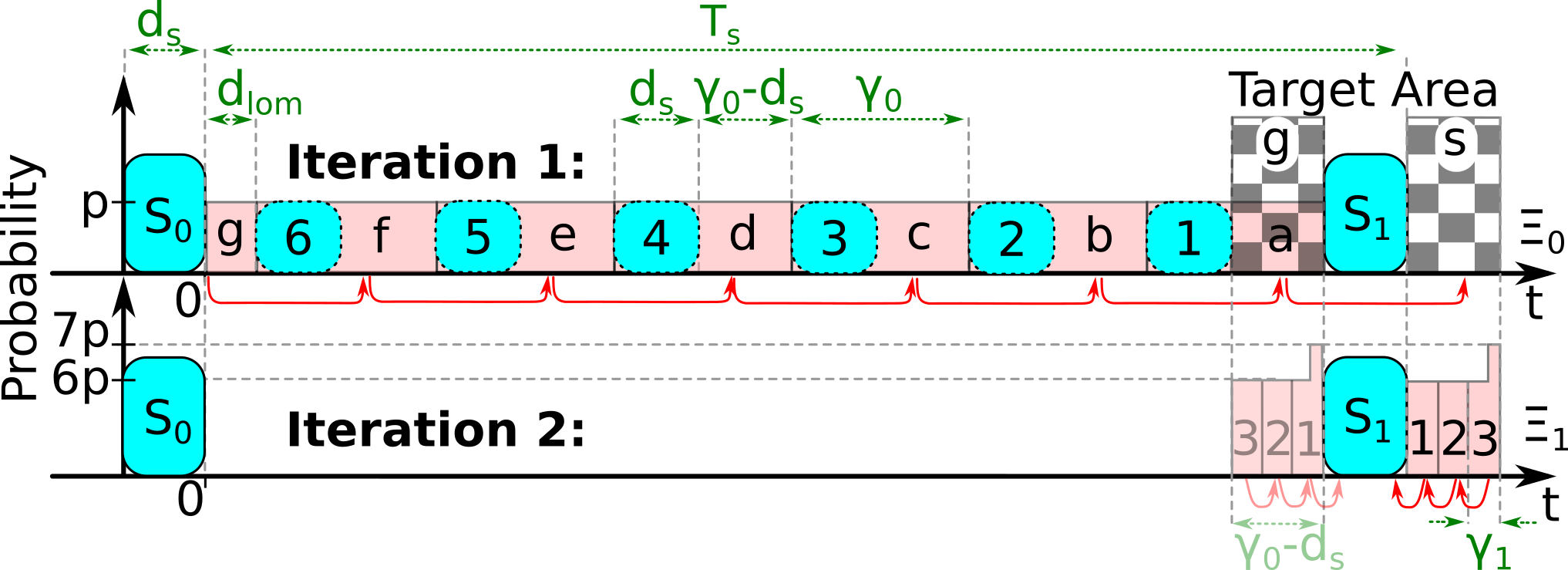}
\caption{Solution scheme of our proposed algorithm.}
\label{fig:algorithm_concept} 
\end{figure}
An example for a neighbor discovery problem that is solved by our proposed algorithm is shown in Figure \ref{fig:algorithm_concept}. As in the previous example from Figure \ref{fig:modelHigherOrderScheme}, there is one growing sequence of order 0 and a shrinking sequence of order 1. 

In \textbf{Iteration 1} of our proposed Algorithm (upper part of Figure \ref{fig:algorithm_concept}), the partitioning for the order-0 sequence, as already described, is carried out. Again, there are \textit{hitting} partitions (1-6 in the figure), from which the scan window on the right ($S_1$) is reached after 1 to 6 steps of length $\gamma_0$. In addition, there are \textit{missing} partitions (a-g), from which the scan window $S_1$ is ``overtaken'' after 1 to 7 steps of length $\gamma_0$.
In this example, each hitting partition has a length of $d_s$, whereas each missing partition has length $\gamma_0 - d_s$, except for partition $g$, which is shortened to fit within $T_s - d_s$.

Since the first advertising packet will fall into every part of $[0, T_s]$ with an equal probability, we assign a probability density of $p = \frac{1}{T_s}$ to every partition that is created in Iteration 1. We can compose the mean discovery latency $\overline{d_{nd}}$ as a sum of multiple probability-weighted latencies ${\sum_k d_{p,k} \cdot p_k \cdot (t_e[k] - t_s[k])} = {\sum_k \overline{d_{p,k}}}$. Here, $d_{p,k}$ is the discovery latency if the initial offset $\Phi[0]$ falls into a certain partition $k$.
We call such a probability-weighted latency a \textit{partial latency} $\overline{d_{p}}$. Each partial latency corresponds to a certain partition type (e.g., to all hitting partitions from Iteration 1, etc.)

In the Example from Figure \ref{fig:algorithm_concept}, all hitting partitions (1-6)	 in Iteration 1 will incur a partial latency of ${\overline{d_{p,0}} = d_s \cdot p \cdot \sum_{k=1}^6 k \cdot \sigma_0}$, with $p = \frac{1}{T_s}$, and the worst-case latency among all hitting parts is $6 \cdot \sigma_0$.
An advertising packet that is sent within a missing partition (a-g) will ``overtake'' the scan window $S_1$ after the corresponding number of advertising intervals. After this overtaking has taken place, the subsequent advertising packet will be sent from one of the two \textit{target areas} depicted in the figure, depending on the mode $m_{n+1}$ of the next higher-order process.  In our example, $m_{n+1} = s$, and the subsequent packet will be sent from target area \textit{s}. This can be represented by ``shifting'' every missing partition by the appropriate multiple of $\gamma_0$ time-units, such that they lie within the target are \textit{s}. Since multiple partitions are shifted into the same area, their probabilities have to be added. For the superposition of the partitions (a, b, c, d, e, f) in Figure \ref{fig:algorithm_concept}, the resulting probability density will be $6 \cdot p$, since each of these partitions has a probability density of $p=\frac{1}{T_s}$. In addition, partition $g$ will increase this merged probability density by $1 \cdot p$ within its corresponding part of the target area, as shown in the figure. The shifting procedure of all missing partitions into the target area \textit{s} incurs a partial latency $\overline{d_{p,1}} = p \cdot \sigma_0 \cdot ((\gamma_0 - d_s) \cdot (1 + 2 + 3 + 5 + 6) + d_{lom} \cdot 7)$. Here, $d_{lom}$ is the length of partition $g$. Similarly, the worst-case latency for a packet being shifted to target area \textit{s} is $7 \sigma_0$. 

The resulting probability distribution within target area \textit{s} is shown in the lower part of Figure \ref{fig:algorithm_concept}, which is represented by the probability buffer $\Xi_1$. 
In \textbf{Iteration 2}, $\Xi_1$ is subdivided into multiple partitions, such that $S_1$ is reached (from target area \textit{s}) after the same multiple of $\gamma_1$ time-units. Again, a partial latency $\overline{d_{p,2}}$ for reaching the scan window from from the subpartitions (1,2,3) of $\Xi_1$ (cf. lower part of Figure \ref{fig:algorithm_concept}) can be computed. 
Here, $\gamma_1 \leq d_s$, and the computation is complete. In other cases, $\Xi_1$ needs to be subpartitioned into hitting and missing parts again, and the procedure repeats.

As can be seen from the Figure, $\Xi_1$ can represent the entire probability distribution after Iteration 1 using only two segments (i.e., one within $[T_s, T_s + \gamma_0 - d_s - d_{lom}]$, and one within $[T_s + \gamma_0 - d_s - d_{lom}, T_s + \gamma_0 - d_s]$). In contrast, in the partitioning scheme described in Section \ref{sec:partitioning}, \textit{every single} missing partition (e.g, all non-hatched partitions in Step 2 in Figure \ref{fig:modelHigherOrderScheme}) need to be processed individually. Hence, this algorithm greatly reduces the computational complexity. 

\subsection{Algorithm Overview}
\begin{figure}[t]
\centering
\includegraphics[width=1.0\columnwidth]{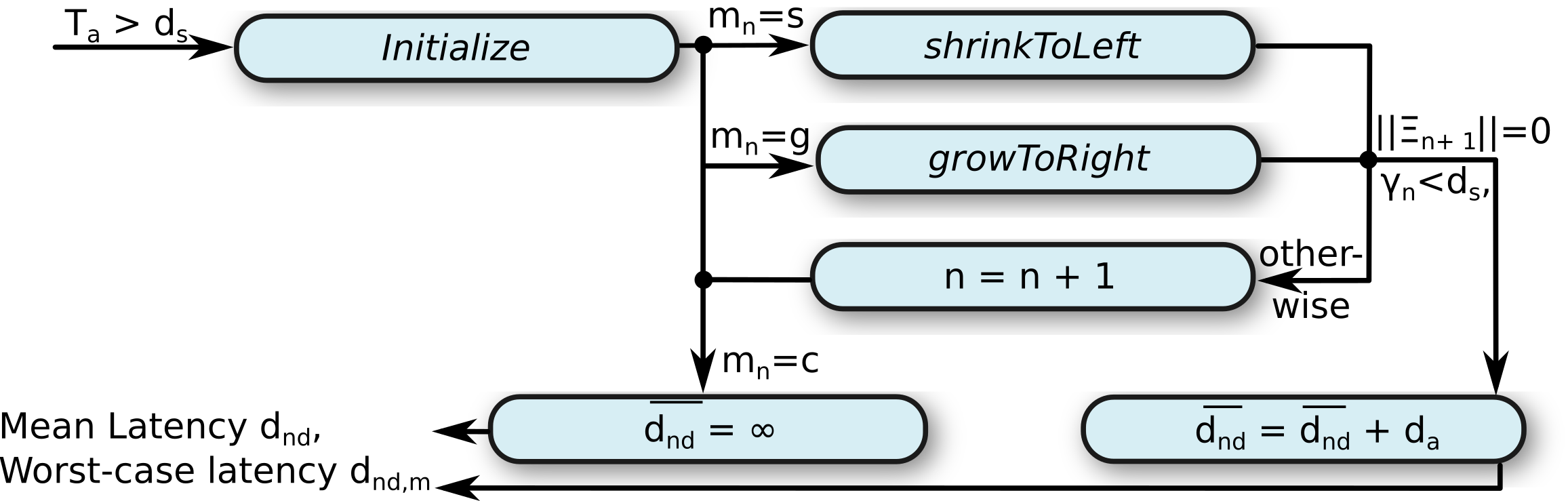}
\caption{Algorithmic solution to assess the discovery latency.}
\label{fig:overall_scheme} 
\vspace*{0.5cm}
\end{figure}

A generic algorithm that corresponds to the described example is shown in Figure \ref{fig:overall_scheme}. 
A formal definition is given in Appendix \ref{app:algorithm}. The algorithm works as follows.

First, the \mbox{\textit{Initialize()}}-function is called, which initializes the mean neighbor discovery latency $\overline{d_{nd}}$ by $0$. Further, an initial probability buffer $\Xi_0$ is created. It contains one segment in $[0, T_s - d_s]$ with a probability of $\frac{1}{T_s}$. The case of the first packet being sent within $[T_s-d_s, d_s]$ is accounted for implicitly by setting the initial probability density to $\frac{1}{T_s}$.

Next, an iterative scheme begins. Depending on the mode $m_n$, either the \textit{growToRight()}- or the \textit{shrinkToLeft()}-function is called. If $m_n = c$, the algorithm is aborted with $\overline{d_{nd}} \gets \infty$, instead.
\textit{growToRight()} performs the partitioning of its input probability distribution and the shifting to the appropriate target area, which results into the new probability buffer $\Xi_{n+1}$ and multiple probability-weighted partial latencies $\overline{d_{p,k}}$. Further, it computes the largest sum of penalties $d_{m}$ for every resulting partition as follows. If a partition to be processed has the largest sum of penalties of $d_{m,1}$, and the highest number of $\gamma$-intervals until reaching the next target area from this partition is $n_i$, then the resulting partition in $\Xi_{n+1}$ will have the largest sum of penalties $d_{m,2} = d_{m,1} + n_i \cdot \sigma_n$ assigned to it. When multiple partitions are merged within one target area, the maximum value of $d_m$ among the merged partitions is always assigned to the resulting partition.
The \textit{shrinkToLeft()}-function provides the equivalent functionality for shrinking sequences. 
After that, one out of two different steps is performed:
\begin{asparaenum}
\item If $\gamma_n < d_s$ or $\Xi_{m+1}$ is empty, then all existing partitions have been examined and $\overline{d_{nd}}$ has been computed. The algorithm terminates. The number of iterations until this occurs is always bounded (cf. Property \ref{prop:limitedOrder}).
\item In all other cases, $n$ is increased and the scheme is repeated by another iteration.
\end{asparaenum}
The mean discovery latency is the sum of weighted partial latencies $\overline{d_{nd}} = E(d)$, and the worst-case latency $d_{nd,m}$ is the largest sum of penalties $d_m$ in the last iteration of the algorithm. The values of $\gamma_n$, $m_{n}$ and $\sigma_n$ are computed at the beginning of every iteration. Finally, since we have assumed the packet length $d_a$ to be $0$, we add $d_a$ to $\overline{d_{nd}}$ to account for the last, successful advertising packet sent. Next, we describe the \textit{growToRight()}-function detail.
%In the next sections, we describe all of the functions used by this algorithm in detail (except \textit{Initialize()}, which has no further functionalities besides the ones already described).

\subsection{\textit{growToRight()}}
\label{sec:gammarization}
Based on a given probability buffer $\Xi_{n}$, which either results from the \textit{Initialize()} - function or from a previous iteration of the algorithm, the \textit{growToRight()}-function iterates through all segments of $\Xi_{n}$ and processes each of them separately. This results into multiple partial latencies $\overline{d_p}$ and one resulting probability buffer $\Xi_{n+1}$.

We first split up each segment of the probability buffer $\Xi_{n}$ into multiple partitions with constant values of $N_{sh}$. $N_{sh}$ denotes the number of steps of length $\gamma_n$ until reaching the right scan window ($S_1$ in Figure \ref{fig:algorithm_concept}) or the appropriate target area (target area s for $m_{n+1} = s$ or target area g for $m_{n+1} = g$), respectively. Next, every such partition is then subdivided into a hitting and a missing part. For every hitting and missing part, a  partial latency $\overline{d_{p}}$ for reaching the scan window is computed. In addition, every missing part contributes to a new segment in $\Xi_{n+1}$, which is obtained by shifting this missing part by $N_{sh} \cdot \gamma_n$ time-units.

If $m_n + 1 = g$, for any given point in time $t \in \Xi_n$, $N_{sh}$ is given by $N_{sh} = \lfloor \frac{t_s - d_s - t}{\gamma_n} \rfloor$.
For $m_{n+1} = s$, the next higher-order sequence shrinks and therefore the advertising packets will approach the temporally right scan window ($S_1$ in Figure \ref{fig:algorithm_concept}) from the right target area (i.e., target area s in Figure \ref{fig:algorithm_concept}).
Further, the temporally right scan window in $\Xi_{n}$ (i.e., $S_1$ in Figure \ref{fig:algorithm_concept}) is regarded as the temporally left one in $\Xi_{n+1}$. This implies that the coordinate system of $\Xi_{n+1}$ has its origin at the end of the temporally right scan window ($S_1$) of $\Xi_n$. Therefore, in addition to shifting every missing partition by $N_{sh} = \lceil \frac{Ts - d_s - t}{\gamma_n} \rceil$ steps of $\gamma_n$ time-units towards the right, the partition has to be transformed into the new coordinate system. The shifting and the coordinate transformation of every point in time $t$ within $\Xi_n$ results into a point in time $t'$ in $\Xi_n{+1}$ of $t' = t - T_s + N_{sh}(t) \cdot \gamma_n$.

In what follows, we formally define the procedure of processing a segment $k \in \Xi_n$. In particular, based on multiple cases, we define the segments that need to be added to $\Xi_{n+1}$ and the partial latencies $\overline{d_{p}}$ need to be summed up to obtain the mean discovery latency.
We first compute the minimum ($N_l$) and maximum number ($N_u$) of $\gamma$-intervals that fit into the distance from every point of this segment to the right scan window, as defined by Equation~\eqref{eq:NlNuGrowing}.
\begin{equation}
\label{eq:NlNuGrowing}
N_l = \left\lceil \frac{T_s - d_s - t_e[k]}{\gamma_n}\right\rceil,\mbox{ }N_u = \left\lfloor\frac{T_s - d_s - t_s[k]}{\gamma_n}\right\rfloor.
\end{equation}

\begin{figure}[htb]
	\centering
	\includegraphics[width=1.0\columnwidth]{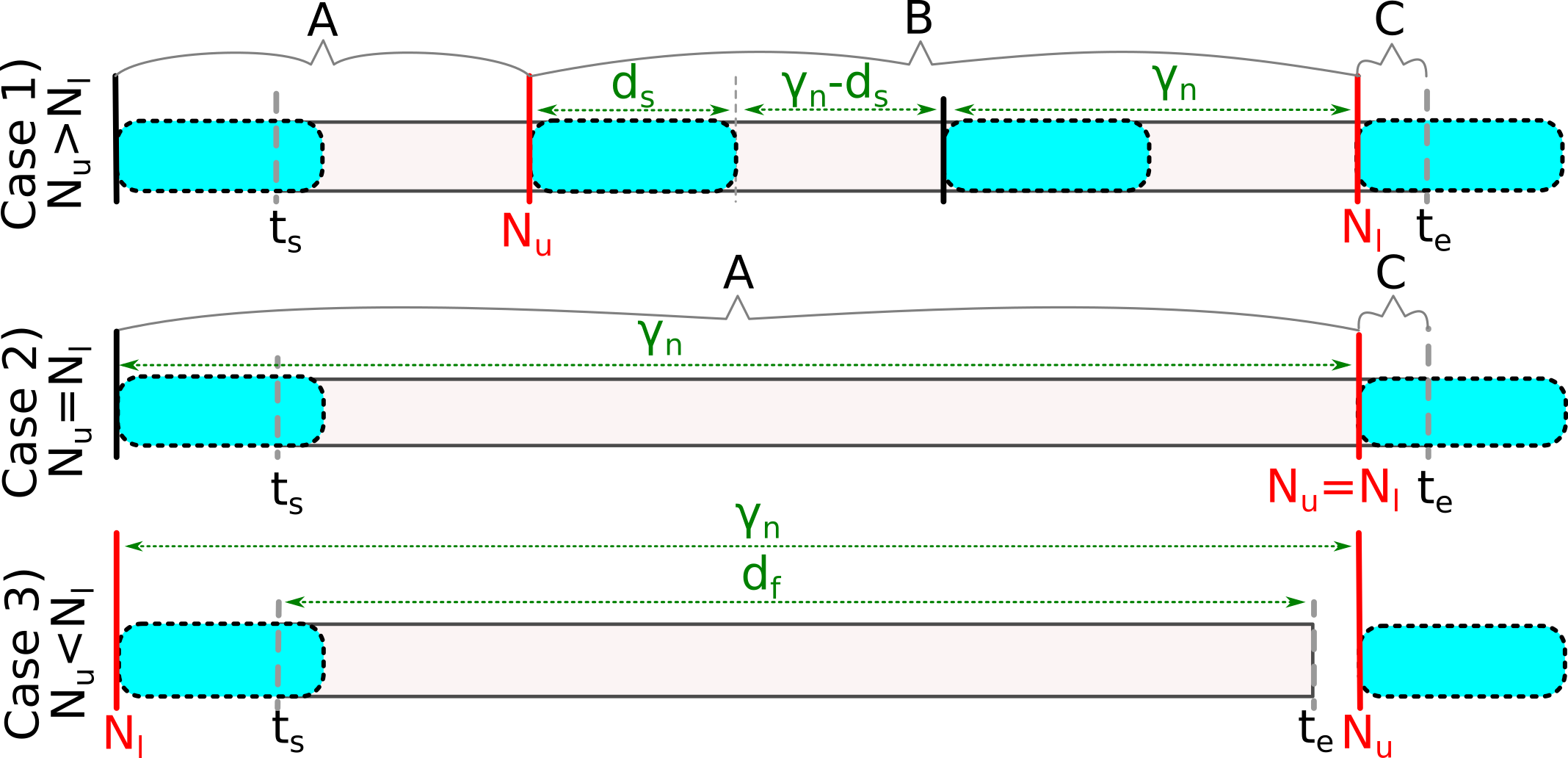}
	\caption{Processing of one probability buffer segment $\Xi[k]$ for growing $\gamma$-sequences}
	\label{fig:gammarization_cases_growing} 
\end{figure}

Figure~\ref{fig:gammarization_cases_growing} depicts a segment of $\Xi_n$ and three possible cases for different values of $N_l$ and $N_u$.
The boxes with dotted rounded borders depict the hitting parts of this segment, whereas the remaining parts represent the missing ones. The figure also shows $N_u$ and $N_l$ and the borders $t_s$ and $t_e$ of the segment. We distinguish between 3 regions in each case: \textbf{A} (probability left of $N_u$), \textbf{B} (probability in between $N_l$ and $N_u$) and \textbf{C} (probability right of $N_l$). The partial latency $\overline{d_{p}}$ of the considered segment $\Xi[k]$ is the sum of all three partial latencies of these regions, viz., $\overline{d_A} + \overline{d_B} + \overline{d_C}$.
%We further define some helping variables by splitting up the segment in its three parts. 
Further, $d_{Nu}$ is the temporal length of the part left of $N_u$, if there is any. Similarly, $d_{Nl}$ is the part right of $N_l$ and $d_f$ is the part in between. These values can be computed as follows.
\begin{equation}
\begin{array}{lcl}
d_{Nu} &=& (T_s - d_s - N_u \gamma_n) - t_s[k].\\
d_f &=& t_e[k] - t_s[k].\\
d_{Nl} &=& t_e[k]  - (T_s - d_s - N_l \gamma_n).\\
\end{array}
\end{equation}

Due to space constraints, we use a shorter notation and write $p$ instead of $\Xi_{n}[k]$, $l$ instead of $t_s[k]$ and $r$ instead of $t_e[k]$. Then, the segments added to $\Xi_{n+1}$ and the partial latencies are as follows.\\

\noindent\textbf{Case 1) and Case 2): $\mathbf{N_u \geq N_l}$}\\
For $m_{n+1} = s$ one has the following partial latencies:
\newcommand{\SetToWidestSecond}[1]{\makebox[\widthof{$(N_u - N_l) (N_u + N_l + 1)$}][l]{$#1$}}
\begin{equation}
\begin{array}{m{0.2cm}m{0.1cm}l}
$\overline{d_{A}}$ &=& p \sigma_n d_{Nu} (N_u + 1).\\
$\overline{d_{B}}$ &=&\left\{\begin{array}{lc} \SetToWidestSecond{0,} & \mbox{if } N_u = N_l, \\ p \sigma_n \gamma_n \frac{1}{2} \cdot & \\\SetToWidestSecond{\cdot (N_u - N_l) (N_u + N_l + 1),}& \mbox{else.}\end{array}\right.\\
$\overline{d_{C}}$ &=& p \sigma_n d_{Nl} N_l.\\
\end{array}
\end{equation}
 For $m_{n+1} = s$, the following segments are added to $\Xi_{n+1}$:
\newcommand{\SetToWidestSecondX}[1]{\makebox[\widthof{${[l - T_s + N_u \gamma_n + \gamma_n, \gamma_n - d_s],} \gets p$}][l]{$#1$}}
\begin{equation}
\begin{array}{l@{\hspace{0.8em}}c@{\hspace{0.8em}}}
\SetToWidestSecondX{[l - T_s + N_u \gamma_n + \gamma_n, \gamma_n - d_s] \gets p,} & \mbox{ if }d_{Nu} < \gamma_n - d_s,\\\SetToWidestSecondX{[0, \gamma_n - d_s] \gets p, }& \mbox{ else.}\\
-,& \mbox{ if } N_u = N_l,\\\SetToWidestSecondX{[0, \gamma_n - d_s] \gets p \cdot (N_u - N_l),}& \mbox{ else.}\\
-,& \mbox{ if } d_{Nl} < d_s,\\\SetToWidestSecondX{[0,r - T_s + N_l \gamma_n]\gets p,}& \mbox{ else.}\\
\end{array}
\end{equation}
For $m_{n+1} = g$, the partial latencies are:
\newcommand{\SetToWidestSecondS}[1]{\makebox[\widthof{$p \sigma_n (N_u - N_l) \cdot (2 \cdot d_s + $}][l]{$#1$}}%
\begin{equation}
\arraycolsep=4.5pt
\begin{array}{m{0.2cm}m{0.1cm}l}
$\overline{d_{A}}$ &=&\left\{\begin{array}{lcl} \SetToWidestSecondS{p \sigma_n d_{Nu} N_u,} & \mbox{if} & d_{Nu} < \gamma_n - d_s, \\ p \sigma_n ((N_u + 1) \cdot& &\\\SetToWidestSecondS{\cdot (d_{Nu} - \gamma_n + d_s)) + }& &\\\SetToWidestSecondS{+ (\gamma_n - d_s) N_u),} & \mbox{else.} & \end{array}\right.\\
$\overline{d_{B}}$ &=&\left\{\begin{array}{lcl} \SetToWidestSecondS{0,} & \mbox{if} & N_u = N_l, \\ \frac{1}{2} p \sigma_n (N_u - N_l) \cdot (2 d_s + & &\\\SetToWidestSecondS{+ \gamma_n\cdot (N_u + N_l - 1)),}& \mbox{else.} &\end{array}\right.\\
$\overline{d_{C}}$ &=&\left\{\begin{array}{lcl} \SetToWidestSecondS{p \sigma_n d_{Nl} N_l,} & \mbox{if} & d_{Nl} < d_s,\\ 
\SetToWidestSecondS{p \sigma_n (d_s + d_{Nl} \cdot (N_l - 1)),} & \mbox{else.} &\end{array}\right.\\
\end{array}
\end{equation}
For $m_{n+1} = g$, the following probabilities are added to $\Xi_{n+1}$: 
\newcommand{\SetToWidestSecondSS}[1]{\makebox[\widthof{${[T_s - \gamma_n, T_s - d_s] \gets p \cdot (N_u - N_l)}$}][l]{$#1$}}%
\begin{equation}
\arraycolsep=4.5pt
\begin{array}{lcl}
[l + N_u \gamma_n, T_s - d_s] \gets p,& \mbox{if} & d_{Nu} < \gamma_n - d_s,\\\SetToWidestSecondSS{[T_s - \gamma_n, T_s - d_s] \gets p,}& \mbox{else.} & \\
-,& \mbox{if} & N_u = N_l,\\\SetToWidestSecondSS{[T_s - \gamma_n, T_s - d_s] \gets p \cdot (N_u - N_l),}& \mbox{else.}& \\
-,& \mbox{if} & d_{Nl} < d_s,\\\SetToWidestSecondSS{[T_s - \gamma_n,r + (N_l - 1) \gamma_n]\gets p,}& \mbox{else.} & \\
\end{array}
\end{equation}
If $\gamma_n < d_s$, no segments are added to $\Xi_{n+1}$ and the partial contribution $\overline{d_p}$ can be simplified (both for $m_{n + 1} = g$ and $m_{n+1} = s$) to $\overline{d_p} = p \sigma_n \cdot (d_{Nl} N_l \frac{1}{2} \gamma_n (N_u - N_l)  \cdot  {(N_u + N_l + 1)} + d_{Nu} (N_u + 1))$.\\
\\
\noindent\textbf{Case 3): $\mathbf{N_u > N_l}$}\\
From Figure \ref{fig:gammarization_cases_growing} follows that this case has three subcases, depending on whether 
$l$, $r$ or both lie within an area for which the current sequence matches, or not. 
For $l - (T_s - d_s - N_l \gamma_N) \leq d_s \land d_{Nl} \leq d_s$,
both $l$ and $r$ lie within a matching area and therefore no probability is added to $\Xi_{n+1}$. The latency $\overline{d_p}$ (both for $m_{n+1} = s$ and $g$) is $p N_l \sigma_n (r - l)$.
For $l - (T_s - d_s - N_l \gamma_N) \leq d_s \land d_{Nl} > d_s$, it is 
\begin{equation}
\overline{d_{p}} = \left\{\begin{array}{lcc} \SetToWidestSecondS{p \sigma_n N_l (r - l),} & \mbox{if} & m_{n+1} = s,\\ 
\SetToWidestSecondS{p \sigma_n (N_l (d_s - \gamma_n + d_{Nu}) + }& &\\\SetToWidestSecondS{ + (N_l - 1)(d_{Nl} - d_s)),}&\mbox{else,} &\end{array}\right.
\end{equation}
and the following segments are added to $\Xi_{n+1}$:

\begin{equation}
\begin{array}{lcc}
{[0, r - T_s - N_l \gamma_n]} \gets p, & \mbox{if} & m_{n+1} = s, \\
{[T_s - \gamma_n, r + (N_l - 1) \gamma_n]} \gets p, & \mbox{if} & m_{n+1} = g .
\end{array}
\end{equation}
Finally, for $l - (T_s - d_s - N_l \gamma_N) > d_s$, it is:
\begin{equation}
\begin{array}{lcl}
\overline{d_p} = p \sigma_n N_l (r - l).& &\\
{[l - T_s + N_l \gamma_n, r - T_s + N_l \gamma_n]} \gets p, & \mbox{if} & m_{n+1} = s, \\
{[l + N_u \gamma_n, r + (N_l - 1) \gamma_n]} \gets p, & \mbox{if} & m_{n+1} = g. 
\end{array}
\end{equation}

If $\gamma_n < d_s$, no segments are added to $\Xi_{n+1}$ and Case 3) can be simplified to $\overline{d_{p}} = p (r - l) N_l \sigma_n$.

The worst-case latency is computed similarly to the mean latency by identifying the largest number of multiples of $\gamma_n$ time-units until reaching either the appropriate target area (for missing partitions) or scan window (for hitting partitions) in each iteration. Sequences with $m_n = s$ are processed by the \textit{shrinkToLeft()}-function, as described in Appendix \ref{app:gammarizeShrinking}.

\section{Analysis and Evaluation}
\label{sec:evaluation}
To evaluate the validity of our proposed theory, we compare its predicted latencies to measured latencies of $29,000$ discovery procedures, which have been carried out by two wireless radios. In addition, to cover a larger fraction of the whole parameter space, we have simulated $81,760,000$ discovery procedures and compare them to the values predicted by our theory.
Towards this, we have carried out 7 different experiments. In Experiment a), we have measured latencies for sweeping advertising intervals and the fixed parameter values \mbox{$T_s = \SI{2.42}{s}, d_s=\SI{0.59}{s}$}. 
In addition, we have simulated the latencies for sweeping values of $T_a$, while $T_s$ and $d_s$ have been set to the following values:
\mbox{b) $T_s = \SI{2.56}{s}, d_s=\SI{0.32}{s}$}; \mbox{c) $T_s = \SI{2.56}{s}, d_s = \SI{0.64}{s}$}; \mbox{d) $T_s = \SI{7.68}{s}, d_s = \SI{0.32}{s}$}; \mbox{e) $T_s = \SI{2.56}{s}, d_s = \SI{0.2015}{s}$},  \mbox{f) $T_s = \SI{5.12}{s}, d_s=\SI{0.32}{s}$}. Further, to study the impact of different scan intervals, we present results obtained from our model for sweeping values of $T_s$ with fixed values \mbox{$T_a = \SI{5.12}{s}, d_s=\SI{0.64}{s}$} in Experiment g) (without comparing them to simulation results). The parameter values for these experiments have been chosen such that they study the impact of different values for $T_a$, $T_s$ and $d_s$ and such that the main characteristics of the latency curves can be examined easily.
In addition, to allow for a quantitative comparison between ground-truth and computed values, we have calculated the maximum and root mean square deviations for these experiments. In particular, the following values are presented, which are based on the simulated or measured mean/max latencies $d_{sim}/d_{mes}$ and the computed latencies $d_{comp}$.
\begin{compactitem}
	\item The Root Mean Square Error $\O$) with
	\begin{equation}
	\O = \frac{\sqrt{\sum_{T_{a,min}}^{T_{a,max}} (d_{comp}(T_a) - d_{sim}(T_a))^2}}{\| T_{a,max - }T_{a,min} \|},
	\end{equation}
	where the denominator is the number of data points with unique advertising intervals. 
	\item The Normalized Root Mean Square Error $\kappa$ with
	\begin{equation}
	\kappa = \frac{RMSE^2}{\max(d_{sim}) - \min(d_{sim})}.
	\end{equation}
	\item The Maximum deviation $MD = \max(\vert d_{comp} - d_{sim} \vert)$.
\end{compactitem}
\vspace*{-0.2cm}
\subsection{Testbed Measurements}
\label{sec:testbed}
To compare the results of our proposed theory against real-world measurements, we conducted the following experiment. Based on the open-source BLE stack \textit{BLESSED} \cite{blessed:15}, we have implemented the protocol studied in this paper. In particular, \textit{BLESSED} implements the BLE protocol, but does not make use of additional random delays added to $T_a$. Advertising and scanning are carried out on one channel. This is exactly the problem described in this paper. We have created two different firmwares for two NRF51822 USB dongles. One device acts as the advertiser. First, it waits for a random amount of time between $[0, T_s]$ and then starts advertising with a certain interval $T_a$. The other firmware scans the channel with an interval $T_s = \SI{2.42}{s}$ and a scan window $d_s = \SI{0.59}{s}$ during all times. For every discovery procedure, the advertiser reports the point in time the advertising is started to a host PC via an USB connection. In the same way, the scanner reports the reception of a packet. With this experiment, the discovery latency can be measured directly. We have conducted the experiment in an anechoic chamber to avoid any interference from external devices (e.g. smartphones). 
To evaluate the accuracy of this setup, we have carried out $10,000$ additional measurements, in which we have measured a maximum loopback latency of $\SI{8}{\milli s}$ between the PC and the wireless node. This means that every measured discovery latency is increased by up to this amount of time.
The advertising interval has been swept between $\SI{0.1}{s}$ and $\SI{3.0}{s}$ in steps of $\SI{0.01}{s}$. Each discovery procedure has been aborted after $\SI{60}{s}$, if no packet has been received within this period. For each advertising interval, the experiment has been repeated $100$ times. The total number of discovery-procedures carried out has been $29,100$, which took about 43 hours of wall-clock time. 
\begin{figure}[b]
	\centering
	\includegraphics[width=\linewidth]{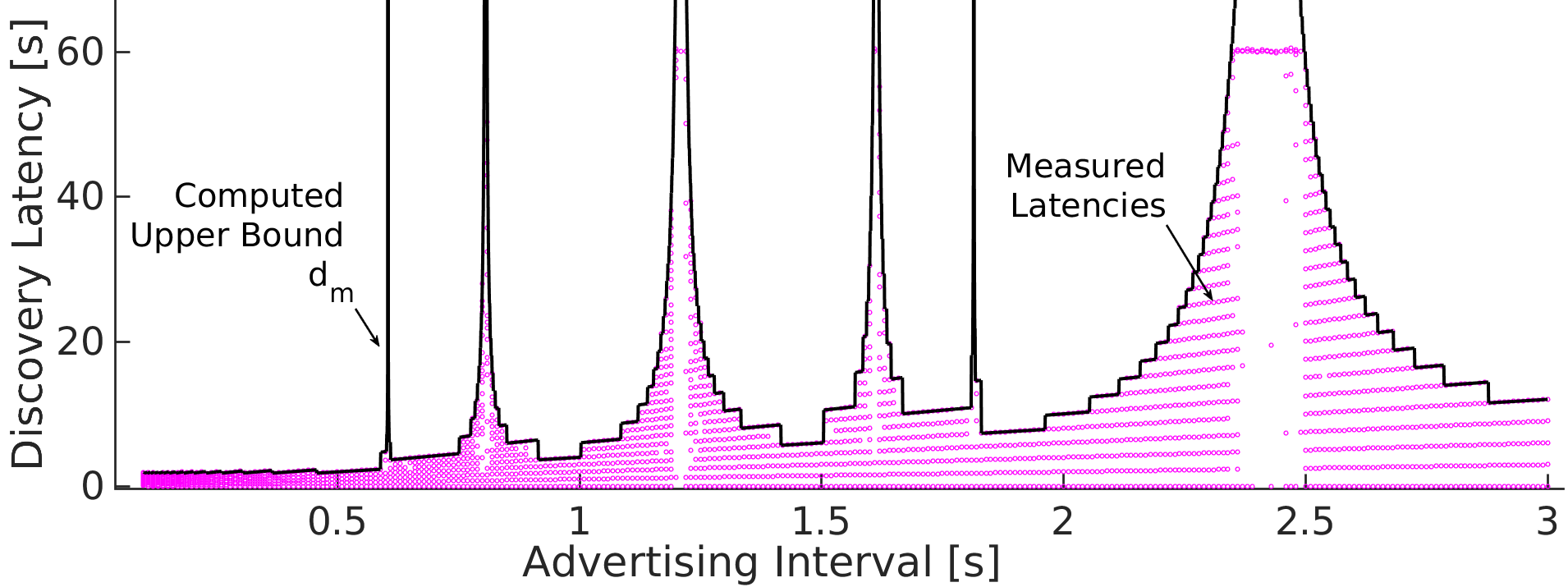}
	\vspace*{-0.5cm}
	\caption{Measured latencies (circles) and computed upper bound (solid line) for Experiment a) ($T_s = 2.42s$ and $d_s = 0.59s$).}
	\label{fig:measurementResults} 
\end{figure}
Figure \ref{fig:measurementResults} depicts the upper latency bound $d_{nd,m}$ predicted by our proposed theory (solid line). In addition, the $29,100$ points in the figure show the measured discovery latencies of the wireless radios. As can be seen, our predicted bound is always reached, but never exceeded. Within latency peaks, most measurements have been aborted before successful discovery was achieved. The abortion took place after $\SI{60}{s}$ of wall-clock time, which explains the accumulation of latencies around $\SI{60}{s}$ within such areas.

To further quantify the accuracy of our theory, among the $100$ repetitions for each advertising interval, the mean and maximum discovery latencies have been determined. 
Figure~\ref{fig:evaluationResults}a) shows the computed mean and worst-case latencies obtained from our theory, together with the measured values for a subset of the advertising intervals that have been realized. Each circle depicts the measured maximum latency out of $100$ discovery procedures, and each cross the mean value. As can be seen, both for the mean and the worst-case latencies, the measurements lie in close proximity with our theory. Mean and maximum errors are given in Row a) of Table \ref{tab:simulation_comparison}. As can be seen, these errors are very small. Values with computed latencies higher than $\SI{60}{s}$ have been excluded from the computation of the errors, since they denote an aborted experiment, as already described. The computed upper latency bound has never been exceeded by more than $\SI{8.7}{\milli s}$ for any of the measured latencies. This error is caused mainly by the latency of the USB connection. The measurements fully confirm the validity of our proposed theory.

\begin{table}[tb]
	\begin{center}
		\caption{Comparison of computation results against real-world measurements (a) and simulations (b-f).}
		\begin{tabular}{|c|c|c|c|c|c|c|}
			\hline  & $\overline{\O}$ & $\overline{\kappa}$ & $\overline{MD}$ & $\O_m$ & $\kappa_m$ & $MD_m$ \\ 
			\hline a) & 0.55 &  0.933 \% & 3.62 & 1.12 & 1.908 \% & 7.32\\  
			\hline b) & 0.49 & 0.06 \% & 6.57 & 0.72 & 0.07 \% & 26.24 \\ 
			\hline c) & 0.35 & 0.05 \% & 6.35 & 0.41 & 0.04 \% & 25.62\\  
			\hline d) & 0.86 &  0.089 \% & 7.21 & 2.61 & 0.26 \% & 114.52 \\ 
			\hline e) & 0.61 &  0.066 \% & 9.02 & 0.66 & 0.067 \% & 38.60 \\ 
			\hline f) & 0.71 &  0.076 \% & 6.90 & 1.88 & 0.19 \% & 83.6 \\ 
			\hline 
		\end{tabular}
		\label{tab:simulation_comparison}
		\vspace*{0.1cm} 
	\end{center}
\end{table}
\vspace*{-0.2cm}
\subsection{Discrete-Event Simulations}
\begin{figure*}[tb]
	\centering
	\includegraphics[width=\linewidth]{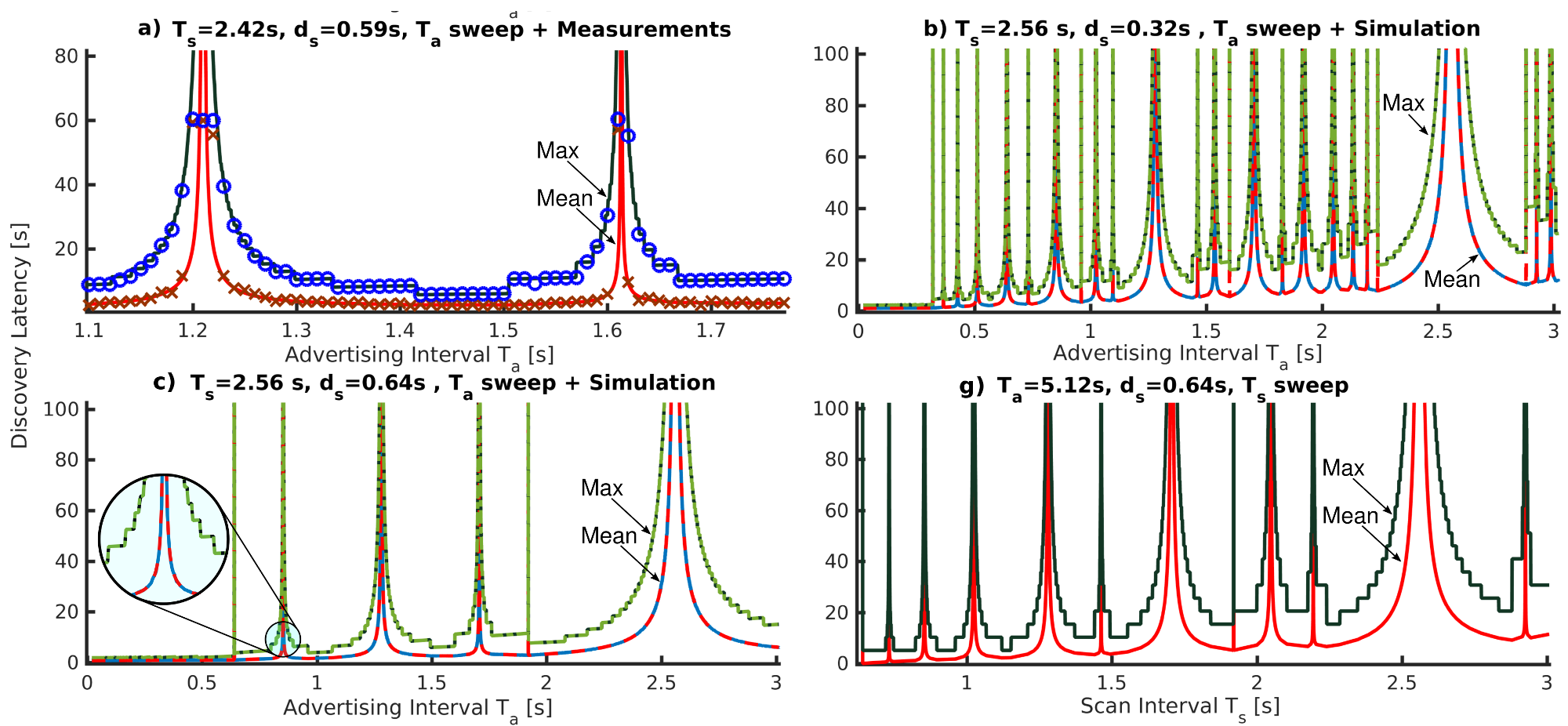}
	\vspace*{-0.5cm}
	\caption{a): Measured (solid line) and computed (dashed line) mean and maximum discovery latencies. b), c): Simulated (solid lines) and computed (dashed lines) mean and maximum discovery latencies. Dashed and solid lines lie in such a close proximity that no difference is visible, indicating a low error between simulated and computed results. Figure g) shows the computed latencies for a fixed value of $T_a$ and varying values of $T_s$.\vspace*{-0.5cm}}
	\label{fig:evaluationResults} 
\end{figure*}
The high amount of wall-clock time required for the measurements limit both the number of repetitions for each advertising interval, as well as the number of examined parameter values. Therefore, we complemented the measurements by comprehensive simulation results. For this purpose, we have developed a custom discrete-event simulator in \mbox{C++}, which carries out the PI-based discovery-process considered in this paper. The initial offset $\Phi[0]$ has been determined by an uniformly distributed random number generator in each simulation. The values of $T_s$ and $d_s$ have remained constant in each of our simulation experiments, while $T_a$ has been varied in steps of $\SI{1.25}{\milli s}$ within $\lbrack \SI{10}{\milli s}\mbox{, }\SI{10.24}{s} - \SI{1.25}{\milli s}\rbrack$.
The packet length $d_a$ has been set to $\SI{248}{\micro s}$, which is a realistic value for protocols such as BLE. To determine the mean latency $\overline{d_{nd}}$ and the maximum latency $d_{nd,m}$, each simulation has been repeated $10,000$ times for every value of $T_a$ and the mean- and maximum latencies have been computed out of the $10,000$ instances. 
The total number of simulations executed for each experiment has been $16,352,000$, which took more than a month of CPU time per experiment. Each single simulation has been aborted after $\SI{1000}{s}$ of simulated time, even if the computed latencies have exceeded this limit.

The Figures \ref{fig:evaluationResults}~b) and c) show the simulated latencies for experiments b) and c) together with the latencies computed by our proposed algorithm for a subset of the advertising intervals considered. In addition, the latencies predicted by our theory for Parametrization g), in which $T_s$ instead of $T_a$ is varied, are shown. The simulated curves are drawn using solid lines, whereas the curves obtained by our algorithm are drawn on top of them using dashed lines. As can be seen, both curves lie in such close proximity that no difference is visible in the figure. The only significant deviations from the simulation results exist at values of $T_a$ at which the maximum latency has exceeded the maximum simulated time. Since each simulation has been aborted at $\SI{1000}{s}$ of simulated time, this behavior is expected.

To carry out a fair numerical comparison, we have excluded all values where the computed maximum latencies lie above $\SI{90}{\percent} \cdot \SI{1000}{s}$ ($\SI{1000}{s}$ is the maximum simulated time). The numerical results of this comparison are given in Rows b)-f) of Table \ref{tab:simulation_comparison}. It can be observed that the root mean square errors $\O$ are negligibly low. This indicates a good match of our proposed theory. Since the maximum predicted latency might not have been reached by any of the $10,000$ simulation repetitions carried out for each value of $T_a$, the maximum errors are higher, in particular for the worst-case latency $MD_m$.
Our predicted upper bound has never been exceeded by any simulated value by more than $\SI{10e-10}{s}$, which we considered as the floating point accuracy. Overall, these results fully confirm our proposed theory.
\vspace*{-0.3cm}
\subsection{Discussion of the Results}
\label{sec:resultDiscussion}

The Figures \ref{fig:evaluationResults}~b) and c) show that the mean discovery latency has a nearly linear relation to the advertising interval for $T_a \leq d_s$. For $T_a > d_s$, both $\overline{d_{nd}}(T_a)$ and $d_{nd,m}(T_a)$ are composed of multiple peaks and minima. If $T_s$ is increased or $d_s$ is decreased, the number of peaks for different values of $T_a$ is increased and these peaks tend to become narrower and steeper. Whereas the mean discovery latency is - except for some singularities - a smooth curve, the figures show (i.e., in the magnifying glass in Figure \ref{fig:evaluationResults}c) ) that the maximum discovery latency is composed of multiple straight lines with points of discontinuity separating them from each other. The situation shown in Figure \ref{fig:evaluationResults}~g), in which $T_s$ is varied, has similar properties. Figure \ref{fig:evaluationResults} reveals interesting results on the behavior of the neighbor discovery latency for different values of $T_a$. First, we can observe that, except for the already mentioned singularities with $m_n = c$, the maximum latency is bounded for all values of $T_a$, $T_s$ and $d_s$. To the best of our knowledge, this fact was previously unknown. It is now clear that interval-based protocols can guarantee deterministic latencies in general. Moreover, it shows that the advertising interval (and hence the duty-cycle) can be greatly increased without increasing the discovery latency by more than a nearly linear relation, if beneficial parametrizations are selected. The highly irregular curves also underpin the need for systematic parameter optimizations in PI-based networks.

Further, new protocols can be developed based on our theory. Such protocols can rely on one or more pairs of optimized advertising and scan intervals. These intervals do not necessarily have to be constant.
As an example, for BLE, strategies that successively reduce the advertising interval (e.g., by half after each interval) have been proposed \cite{liu:12_long} and can now be optimized systematically using our theory.

Using our proposed solution, unsynchronized continuous data transmission with duty-cycled, PI-based protocols (also referred to as continuous broadcasting, e.g. in the context of BLE beacons) becomes feasible, since the broadcaster and the receivers can be configured such that the number of packet repetitions is optimized and a deterministic reception is guaranteed. Since it is now known after how many packets a reception takes place in the worst-case, finding optimal configurations is straight-forward.

\vspace*{-0.3cm}
\subsection{Comparison to Existing Models}
\label{sec:model_comparsion}

\begin{figure}[b]
	\centering
	\includegraphics[width=\linewidth]{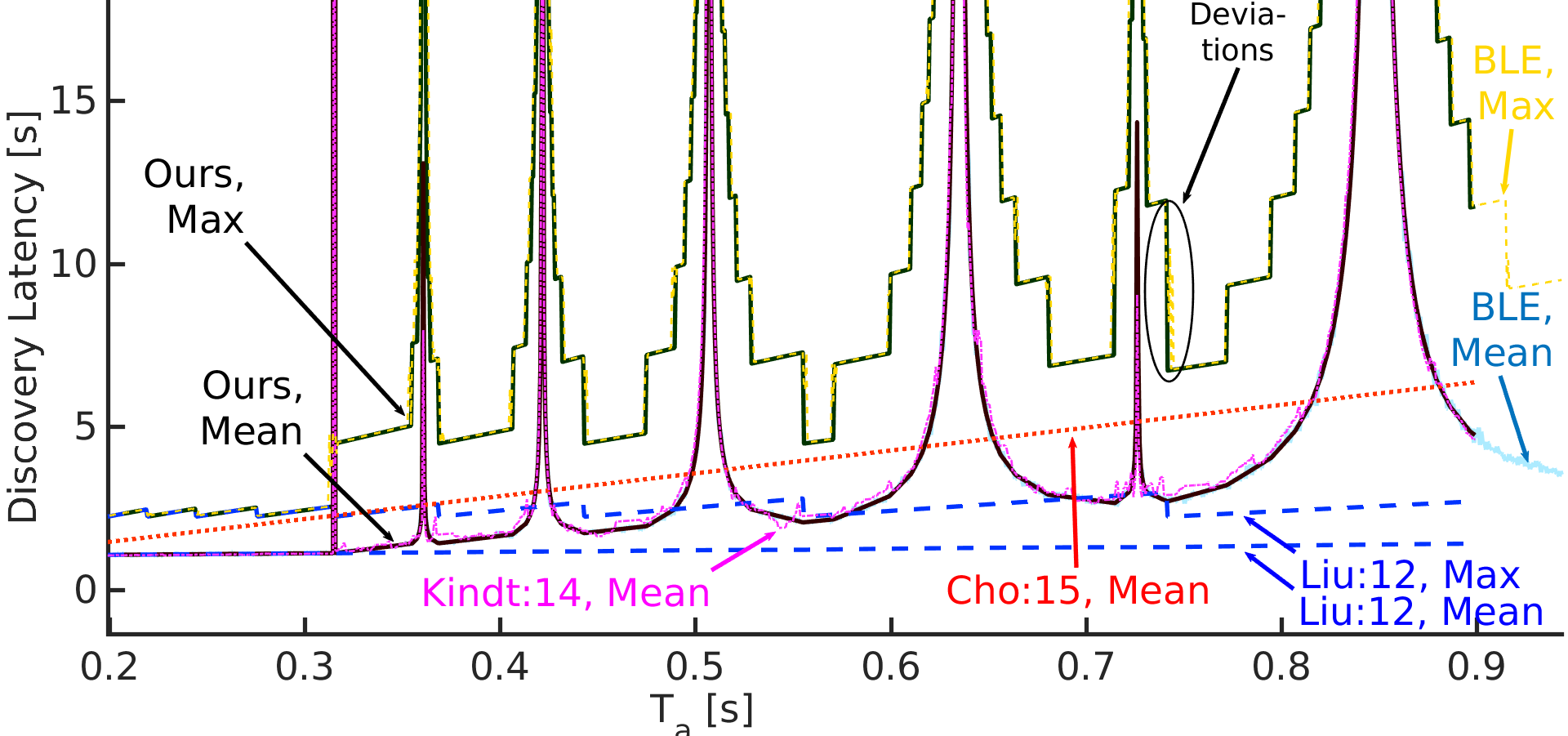}
	\vspace*{-0.5cm}
	\caption{Comparison to the models Liu:12 \cite{liu:12_short}, \cite{liu:12_long}, \cite{liu:12_techrep}, \cite{schurgers:02}, Cho:15 (\cite{cho:15}, \cite{cho:15_2}), Kindt:13 \cite{kindt:13}, our proposed model and a simulation of BLE.}
	\label{fig:modelComparison} 
\end{figure}

In this  section, we compare our proposed theory to existing models for BLE. We assume the same parametrization as for case b). For a fair comparison, we have implemented modified versions of these models, such that they describe the ideal PI-based problem, rather than the 3-channel discovery procedure of BLE.
We have considered the model by Liu et al. \cite{liu:12_short}, \cite{liu:12_long}, \cite{liu:12_techrep}, which is identical to the model from \cite{schurgers:02} when considering single-channel discovery. Besides our proposed theory, this model is the only one which predicts upper bounds (for $T_a < d_s$, only). Further, we have considered the predicted mean latencies of the model by Cho et al. \cite{cho:15}, \cite{cho:15_2} and the accelerated simulation model by Kindt et al. \cite{kindt:13}, with $\Delta = \SI{93.12}{s}$ and $\epsilon = 0.99$ into this comparison. 

To compare to which extent existing models approximate the BLE protocol, we have conducted discrete-event simulations of the complete BLE discovery procedure, with 1000 repetitions per value of $T_a$. For this purpose, the simulator used for the evaluation of our model has been extended to account for 1) three channel discovery and 2) a random delay $\rho \in [\SI{0}{s},\SI{10}{ms}]$ being added to each advertising interval.
As already described, the outputs of the simulator for the ideal PI discovery procedure are in line with the results obtained by our model, which are verified against real-world measurements (cf. Section \ref{sec:testbed}).
We have assumed an effective advertising interval $T_a' = T_a + \SI{0.005}{s}$ to account for the mean shift of BLE's random delay. This increases the accuracy of all models considered.

\noindent \textbf{Comparison with other models:} As shown in Figure \ref{fig:modelComparison}, the model by Liu \cite{liu:12_short}, \cite{liu:12_long}, \cite{liu:12_techrep}, \cite{schurgers:02} matches well for $T_a < d_s$, but is not valid for $T_a > d_s$.  It can be seen from the figure that the model by Cho \cite{cho:15}, \cite{cho:15_2} is unable to predict any latency peaks, due to the assumption of independent overlap probabilities. The simulation model by Kindt \cite{kindt:13} models the overall trend, but is subjected to significant statistical variations.

\noindent  \textbf{Comparison with BLE}: The results of this comparison are shown in Figure \ref{fig:modelComparison}. As can be seen, our proposed model approximates the properties of BLE well. Small deviations of the predicted upper latency bound at border values can be observed e.g. around $T_a \approx 0.74s$, which are caused by the random delay and the multichannel discovery. While our model provides, to the best of or knowledge, the most accurate latency estimations for the BLE protocol that are available, it needs to be mentioned that for some parametrizations, the random delay could cause significant deviations from our predicted latencies. Studying the random delay and three-channel discovery is left open for future research.
\vspace*{-0.2cm}
\subsection{Computational Complexity}
\label{sec:compComplex}
In this section, we evaluate the computational complexity of Algorithm \ref{alg:algorithmDefintion}. Our aim is to establish that it has reasonable computational overhead and can therefore be used to compute a large number of latency values in short amounts of time. 

The algorithm processes each segment of a probability buffer $\Xi_n$, evaluates the corresponding equations and adds new probability segments to the next probability buffer $\Xi_{n+1}$. The computational complexity for processing any order $n$ is therefore determined by the number of segments in $\Xi_n$. Whereas the initial probability buffer contains only one segment, a conservative assumption is that each additional sequence-order $n \gets n+1$ potentially triples the number of entries in $\Xi_{n+1}$, since each segment of $\Xi_n$ might be subdivided twice in the worst-case. Therefore, the complexity grows exponentially with larger maximum orders $n_m$. However, the maximum order is bounded, and the bound grows only logarithmically with the smaller interval of $T_a$ and $T_s$, as defined by Equation~\eqref{eq:maxIterationOrder}. Because practical problems (e.g., BLE) have limited maximum orders, the expected computational demands are low.

\begin{figure}[htb]
\centering
\includegraphics[width=\linewidth]{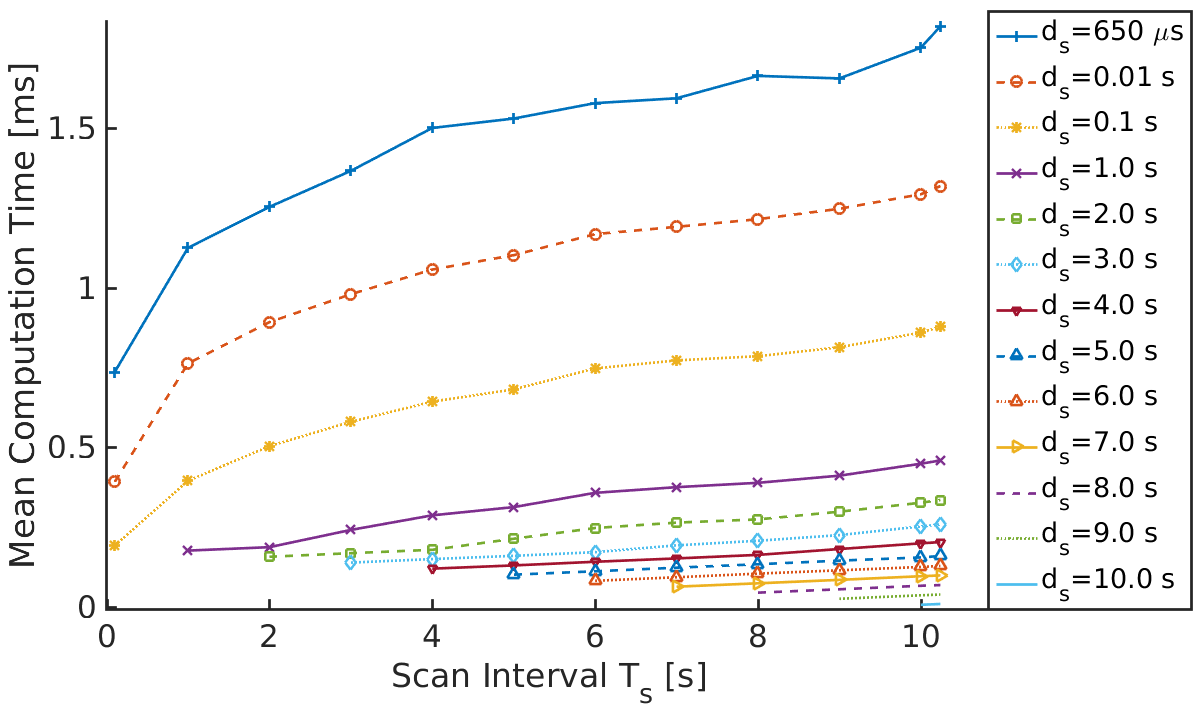}
\caption{Mean computation time per advertising interval for computing the mean- and maximum latencies for all advertising intervals $T_a \in [\SI{0.2}{s}\mbox{, }\SI{10.24}{s}]$ in steps of $\SI{650}{\micro s}$.\vspace*{-2.0em}}
\label{fig:computationalComplexity} 
\end{figure}

For a detailed evaluation, we have measured the computation time of a MATLAB-implementation using the tic/toc profiling mechanism of MATLAB \cite{matlabTic:15}. The timing measurements were taken on a Lenovo Thinkpad X240 laptop with MATLAB R2015b under Linux Mint 17.2, using a single CPU core. For different values of $d_s$ and $T_s$, we ran our proposed algorithm for all advertising intervals between $[\SI{20}{\milli s}, \SI{10.24}{s}]$ in steps of $\SI{650}{\micro s}$, which is equivalent to the whole range of advertising intervals allowed in BLE. For each of these $16,353$ executions of Algorithm \ref{alg:algorithmDefintion}, the execution time was measured. The mean time the algorithm took until termination among the $16,535$ executions is shown in Figure~\ref{fig:computationalComplexity} for multiple scan intervals and windows. The plot also includes the extreme case for BLE, with $d_s = \SI{650}{\micro s}$ and $T_s = \SI{10.24}{s}$. This parametrization leads to the maximum theoretically possible order of $n_m = 14$ for BLE\footnote{During this experiment, the maximum problem order that actually appeared was 10.}. 
The highest computation time reached has been $\SI{1.8}{\milli s}$ for $d_{s} = \SI{650}{\micro s}$ and $T_s = \SI{10.24}{s}$. The results show that our proposed algorithm is computationally cheap in real-world situations and multiple orders of magnitude faster than discrete-event simulations. Recall that for the curves in Figure \ref{fig:evaluationResults}, more than a month of computation time has been spent for each simulation.

\section{Parameter Selection}
\label{sec:parameter_selection}
In this section, we briefly describe how our proposed theory can be used to optimize the parametrization of PI-based discovery approaches.
As shown in Section \ref{sec:compComplex}, Algorithm \ref{alg:algorithmDefintion} is computationally cheap and therefore well suited to be executed repeatedly for a large set of parameters. For example, for a fixed tuple of values $(T_s$, $d_s)$, all possible values of $T_a$ allowed by the BLE specification \cite{bleSpec} can be computed within a few seconds. With simulations, this is not possible due to the long simulation times and the limited accuracies. Therefore, for optimizing the parameters, we propose performing a design-space exploration based on our model. The optimization goal is specific to the application and is typically one out of the following possibilities:
\begin{asparaenum}
\item Minimize the mean/maximum discovery latency as the first optimization goal. Since the same latencies can be achieved with multiple parametrizations, a second optimization criterion can be chosen.
\item Minimize the expected mean/maximum energy-consumption spent for the discovery. Given the energy $E_a$ for an advertising packet and $E_s$ for one scan window, a good approximation of this energy is $\overline{E_{nd,a}} = E_a \cdot \frac{\overline{d_{nd}}}{T_a} d_a$ for the advertiser and $\overline{E_{nd,s}} = E_{s} \cdot \frac{\overline{d_{nd}}}{T_s} d_s$ for the scanner.
When considering the maximum discovery latency $d_{nd,m}$ instead of $\overline{d_{nd}}$ and by setting $E_a = E_s = 1$, this equation approximates the power-latency-product defined in \cite{Kandhalu:10}.
\item Minimize the joint energy consumption of both the advertiser and the scanner, such that their sum becomes minimal.
\end{asparaenum}

\begin{figure}[tbh]
\centering
\includegraphics[width=1.0\columnwidth]{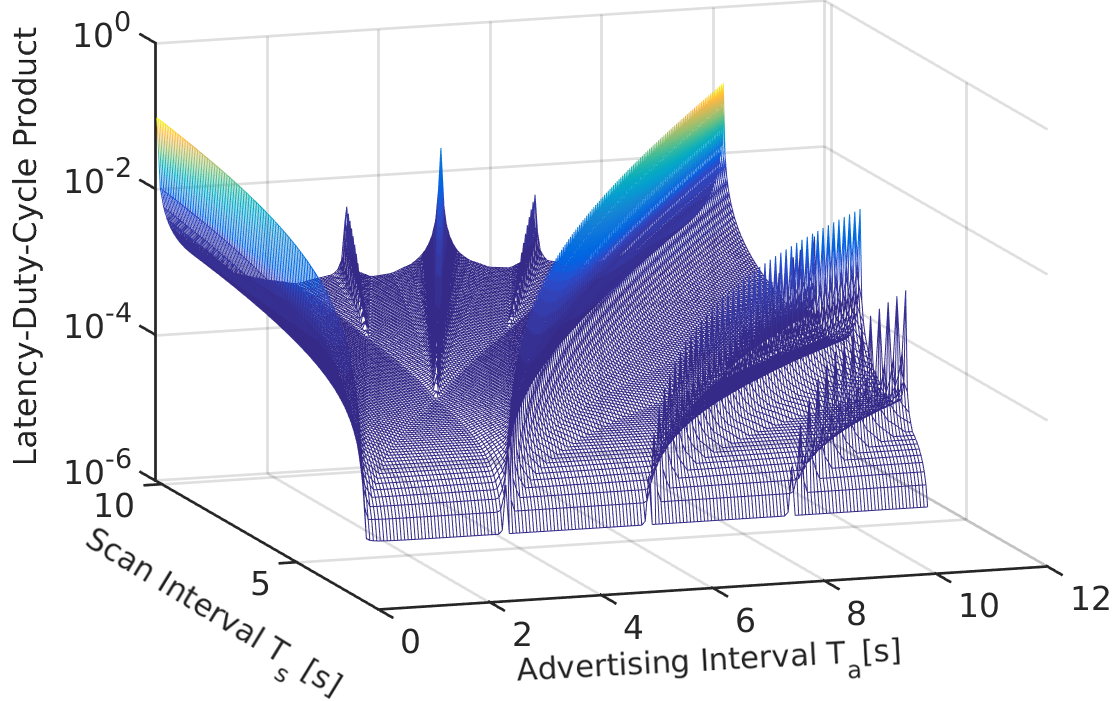}
\caption{Latency-duty-cycle product for a fixed scan window $d_s = 2.5 s$.}
\label{fig:meshPlot} 
\end{figure}
A design space exploration for minimizing the mean-power-latency product is shown in Figure \ref{fig:meshPlot}. For a fixed scan window $d_s = 2.5s$, $T_a$ and $T_s$ have been varied in steps of $62.5 ms$ to compute the latency-duty-cycle product $\frac{d_a}{T_a} \cdot \overline{d_{nd}}$ of the advertiser. In the figure, all latencies exceeding $\SI{1e10}{s}$ have been truncated to this value. As can be seen, there are multiple optimal points for a given scan interval, such that the advertising interval can be chosen within a large set of values without affecting the energy efficiency. 
\section{Concluding Remarks}
\label{sec:conclusion}
We have presented a new theory for computing the neighbor discovery latencies of periodic interval-based, slotless protocols. Our results imply two major novel insights, which have important implications. 

First, the discovery latency of such protocols can now for the first time be computed accurately. Since PI-based concepts are widely used (e.g., for the BLE protocol and for ANT/ANT+), these protocols can now be configured in an optimized manner. Until now, the specifications of some wireless services recommend certain parameter ranges in which latency-peaks can occur. For example, the parametrization suggested by the BLE find-me profile \cite{BleFindMeProf} allows interval lengths that lie within such a peak, leading to long discovery latencies \cite{kindt:13}. Therefore, such recommended parametrizations should be derived using our theory in the future.

Second, our results have revealed that the latency is bounded for almost all parameter values. This makes PI-based solutions applicable to networks with deterministic latency demands, where currently slotted protocols are being used. Periodic interval-based solutions provide the largest flexibility among all deterministic possibilities. Recent work based on this paper \cite{kindt:16}, \cite{kindt:17a} has revealed that with the same duty-cycle, PI-based protocols can achieve significantly shorter worst-case latencies.
Also, novel protocols can be developed based on our theory, as described in Section \ref{sec:resultDiscussion}. Our next step will be adopting the theory to commonly used protocols (such as BLE) by accounting for their unique peculiarities, such as multiple channels and random delays. 
Finally, with our results, we hope to motivate other researchers to work on slotless neighbor discovery protocols. 

\section*{Acknowledgments}
The authors would like to thank the anonymous reviewers and the editors for their constructive feedback on the earlier versions of this paper.
This work was partially supported by \textit{HE2mT - High-Level Development Methods for Energy-Saving, Mobile Telemonitoring Systems}, a project funded by the federal ministry of education and research of Germany (BMBF).

	\pagebreak
	\bibliographystyle{IEEEtran}
	\bibliography{literature}

	\begin{appendix}
\section{Recursive Computation of $\gamma_n$}
\label{app:recursiveGammaComputation}
In Section \ref{sec:computingGamma}, we have described how to compute the value of the parameter $\gamma_n$. For sequences with orders $n > 1$, a recursive scheme to compute $\gamma_{n}$ given $\gamma_{n-1}$ has been presented for two consecutive growing sequences. Though it has been claimed that the equations presented also hold true for two consecutive shrinking processes, this case has not been explained. In this appendix, we first explain two consecutive shrinking sequences. Then, we describe recursion schemes for sequences with $m_n \ne m_{n+1}$. Again, the multiple $Q_n$ is given by $Q_n = \lfloor \frac{d_{t,n}}{\gamma_{n}} \rfloor$.\\
\\
\textbf{Shrinking} $\rightarrow$ \textbf{Shrinking:}\\
\begin{figure}[htb]
	\centering
	\includegraphics[width = 1.0\columnwidth]{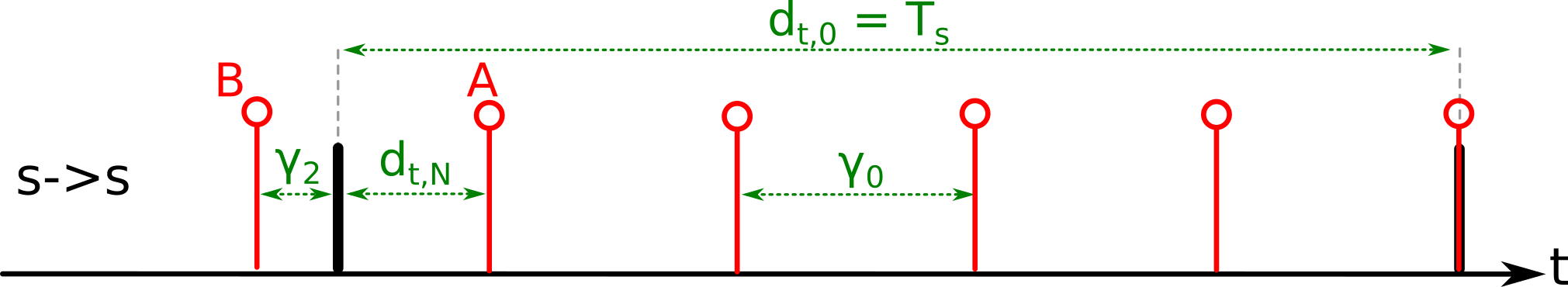}
	\vspace*{-0.5cm}
	\caption{Computation of $\gamma_1$ given $\gamma_0$ for $m_0 = m_1 = s$.\vspace*{-0.5cm}}
	\label{fig:computegamma_shsh} 
\end{figure}

\begin{figure}[htb]
	\centering
	\includegraphics[width = 1.0\columnwidth]{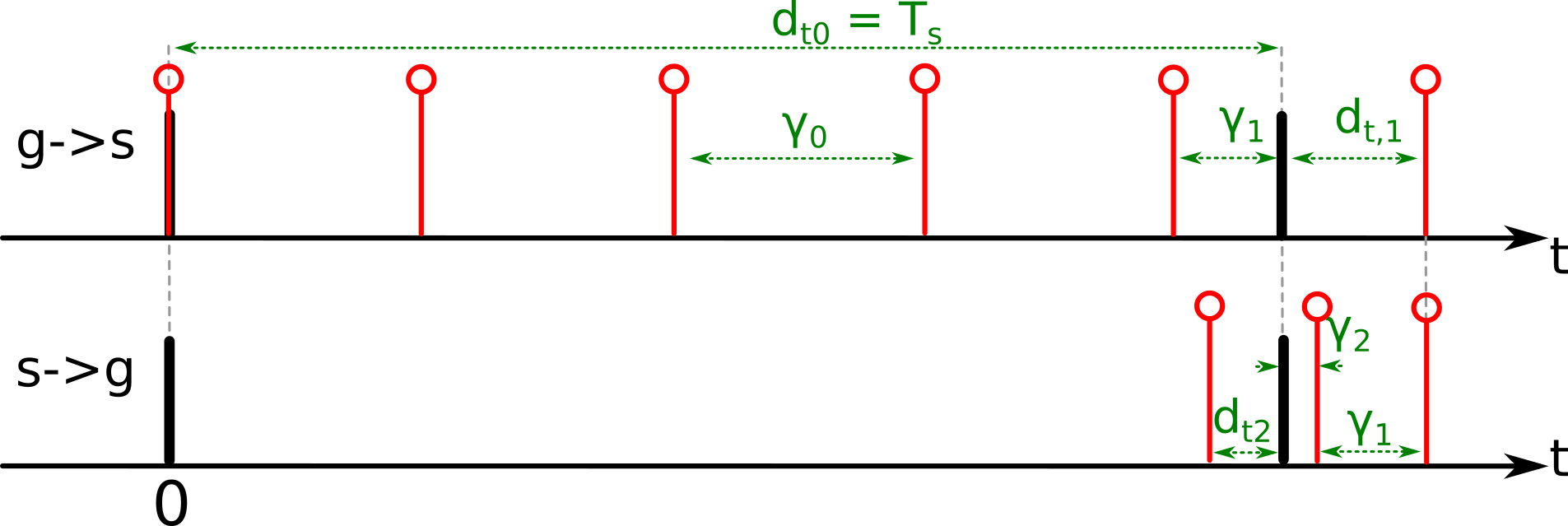}
	\vspace*{-0.5cm}
	\caption{Computation of $\gamma_1$ given $\gamma_0$ for $m_0 \ne m_1 \ne m_2$.\vspace*{-0.5cm}}
	\label{fig:computegamma_grsh} 
\end{figure}

A situation with $m_{0} = m_1 = s$ is shown in Figure \ref{fig:computegamma_grsh}. For $m_1 = m_2 = g$, we have formed the linear combination with the smallest possible sum $\sum_{k=0}^{n-1} a_k \cdot \sigma_k$ that exceeds $T_s$ by no more than $\frac{1}{2} \gamma_0$. Here, since $m_0 = s$, the coefficient $a_0$ of the linear combination is negative and hence we form the negative combination of $\gamma_0$ time-units for which the absolute value exceeds $T_s$ by no more than $\frac{1}{2}\gamma_0$ time-units. This also leads to Equation \ref{eq:gammaNextGrGr}.

\textbf{Growing} $\rightarrow$ \textbf{Shrinking and Shrinking} $\rightarrow$ \textbf{Growing:}\\

Figure \ref{fig:computegamma_grsh} depicts a situation in which $m_0 = g$, $m_1 = s$ and $m_2 = g$. When comparing this to the situation in Figures \ref{fig:computegamma_grgr} and \ref{fig:computegamma_shsh}, it becomes clear that when computing $\gamma_n$, a mode-change $m_{n-1} \ne m_{n}$ incurs one instance of $\gamma_{n-1}$ fewer than for $m_n = m_{n-1}$. This leads to the following Equations, which hold true for $m_n = s$, $m_{n+1} = g$ and $m_{n} = g$, $m_{n+1} = s$. 

\begin{equation}
\label{eq:gammaNextShGr}
\begin{array}{lcl}
\gamma_n		&	=	&	d_{t,n-1}  - Q_{n-1} \cdot \gamma_{n-1}.\\
d_{t,n}		&	=	&	(Q_{n-1} + 1) \cdot \gamma_{n-1} - d_{t,n-1}.\\
\sigma_n		&	=	&	\sigma_{s,n-1} + Q_{n-1} \cdot \sigma_{n-1}.\\
\sigma_{s,n}	&	=	&	\sigma_{s,n-1} + (Q_{n-1} + 1) \cdot \sigma_{n-1}.
\end{array}
\end{equation}

\section{\textit{shrinkToLeft()} - function}
\label{app:gammarizeShrinking}
In Section \ref{sec:gammarization}, we have presented the \textit{growToRight()}-function, which grows a given distance $\Phi$ towards the temporally right scan window. Whereas \textit{growToRight()} only works for growing sequences (i.e., $m_n = g$), \textit{shrinkToLeft()} does the equivalent computation for shrinking sequences with $m_n = s$ of any order $n$, thereby reducing $\Phi$ ins steps of $\gamma_n$ towards the temporally left scan window.
\begin{figure}[htb]
	\centering
	\includegraphics[width=1.0\columnwidth]{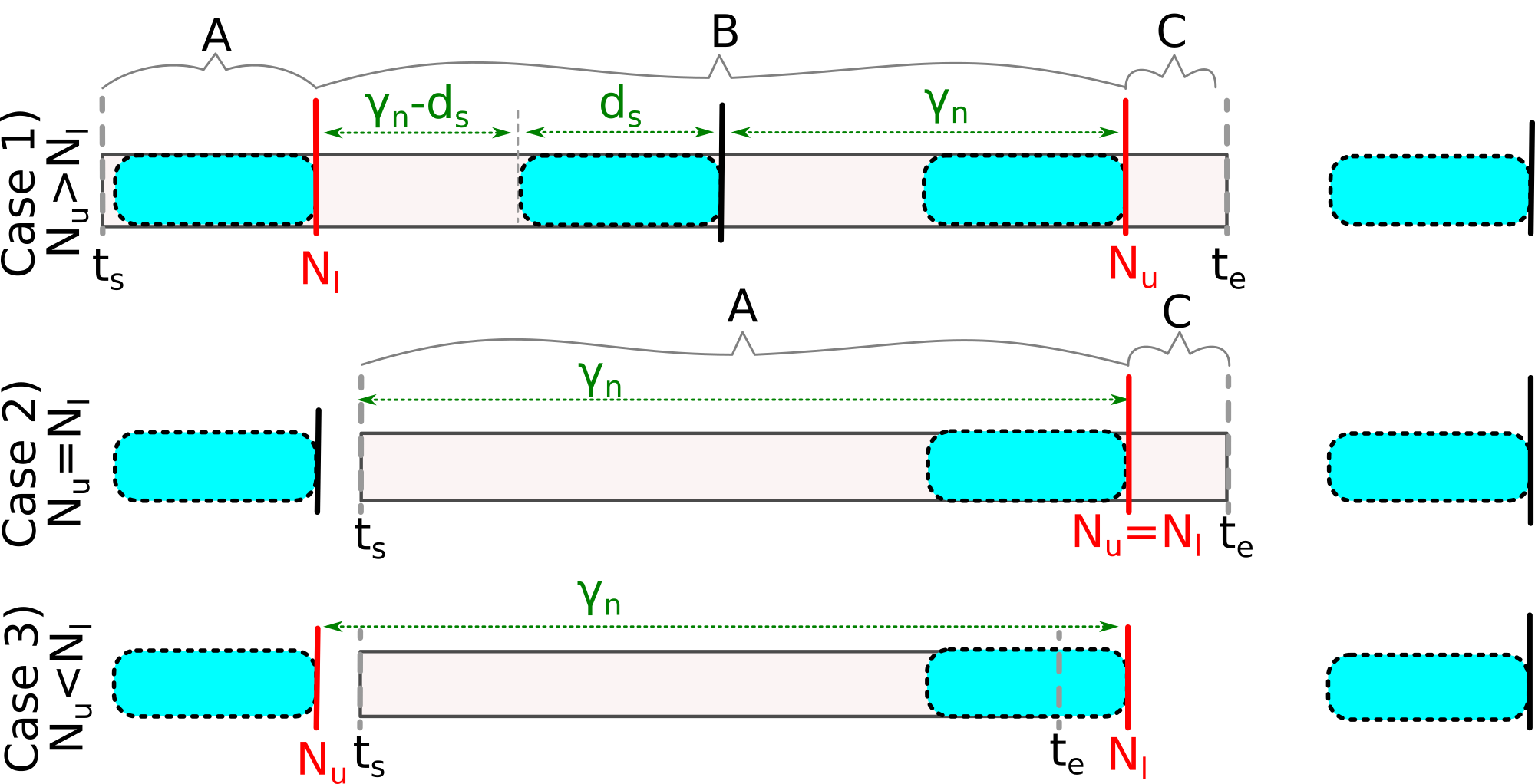}
	\caption{Processing of one probability buffer segment $\Xi[k]$ for shrinking $\gamma$-sequences.}
	\label{fig:gammarization_cases_shrinking} 
\end{figure}

The partitioning of a segment for a shrinking sequence is depicted in Figure \ref{fig:gammarization_cases_shrinking}. The derivation of the equations is similar to the \textit{growToRight()}-function and hence we only present the final results without further explanations.
$N_u$ and $N_l$ are defined as
\begin{equation}
\begin{array}{lr}
N_l = \left\lceil \frac{t_s[k]}{\gamma_n} \right\rceil, & N_u = \left\lfloor \frac{t_e[k]}{\gamma_n} \right\rfloor.
\end{array}
\end{equation}
Due to space constraints, we use a shorter notation and write $p = \Xi_{n}[k]$, $l$ instead of $t_s[k]$ and $r$ instead of $t_e[k]$.
The variables $d_{Nl}$, $d_{Nu}$ and $d_f$ are, in analogy to \textit{growToRight()}:
\begin{equation}
\begin{array}{lcr}
d_{Nl} = N_l \gamma_n - l, & d_{Nu} = r - N_u \gamma_n, & d_f = r - l.\\
\end{array}
\end{equation}
Like in \textit{growToRight()}, there are three different cases.\\
\\
\textbf{Case 1) and Case 2): $\mathbf{N_u \geq N_l}$}\\
For $m_{n+1} = s$, the partial latencies of the three parts \textbf{A}, \textbf{B} and \textbf{C} as depicted in Figure \ref{fig:gammarization_cases_shrinking} are defined as follows.

\newcommand{\SetToWidestGShift}[1]{\makebox[\widthof{$+ (d_{Nu} - \gamma_n + d_s) (N_u a))$}][l]{$#1$}}%
\begin{equation}
\arraycolsep=4.5pt
\begin{array}{l@{\hspace{0.1em}}l@{\hspace{0.1em}}l@{\hspace{0.1em}}}
\overline{d_{A}}  &=& \left\{\begin{array}{lcl} \SetToWidestGShift{p \sigma_n N_l d_{Nl},}& \mbox{if} & d_{Nl} < d_s, \\ p \sigma_n (d_s N_l + & &\\ + (d_{Nl} - d_s)(N_l - 1)),& \mbox{else.} & \end{array}\right.\\
\overline{d_{B}} &=&\left\{\begin{array}{lcl} \SetToWidestGShift{0,} & \mbox{if} & N_u = N_l, \\ \frac{1}{2} p \sigma_n (N_u - N_l) \cdot & &\\\SetToWidestGShift{\cdot(2 d_s + \gamma_n (N_u + N_l - 1)), }& \mbox{else.}& \end{array}\right.\\
\overline{d_{C}} &=&\left\{\begin{array}{lcl} p \sigma_n N_u d_{Nu}, & \mbox{if} & N_l \gamma_n - r \geq d_s, \\ p \sigma_n ((d_{Nl}- d_s) (N_l-1) +  & &\\\SetToWidestGShift{+ (d_s - \gamma_n + d_{Nu}) N_l)},& \mbox{else.}& \end{array}\right.\\
\end{array}
\end{equation}
The following probability segments are added to $\Xi_{n+1}$ for $m_{n + 1} = s$. 
\newcommand{\SetToWidestZShift}[1]{\makebox[\widthof{$+ (d_{Nu} - \gamma_n + d_s) (N_u aaaaaaaa))$}][l]{$#1$}}%
\begin{equation}
\begin{array}{l@{\hspace{0.8em}}c@{\hspace{0.8em}}l@{\hspace{0.8em}}}
\SetToWidestZShift{-,} & \mbox{if} & d_{Nl} < d_s, \\
\SetToWidestZShift{{[\gamma_n - d_{Nl}, \gamma_n - d_s]} \gets p,}&\mbox{else.}& \\
\SetToWidestZShift{-,}& \mbox{if} & N_u = N_l,\\{[0, \gamma_n - d_s]} \gets p \cdot (N_u - N_l),& \mbox{else.}\\
\SetToWidestZShift{{[0, d_{Nu}]} \gets p,}& \mbox{if} & d_{Nu} < \gamma_n - d_s,\\{[0,\gamma_n - d_s]}\gets p,& \mbox{else.}
\end{array}
\end{equation}

For $m_{n+1} = g$, the following partial latencies $\overline{d_p}$ exist:
\newcommand{\SetToWidestGAAShift}[1]{\makebox[\widthof{$+ (d_{Nu} - \gamma_n + d_s) (N_u$}][l]{$#1$}}%
\begin{equation}
\begin{array}{l@{\hspace{0.1em}}c@{\hspace{0.1em}}l@{\hspace{0.1em}}}
\overline{d_{A}}  &=& \SetToWidestGAAShift{p \sigma_n N_l d_{Nl}.} \\
\overline{d_{B}} &=&\left\{\begin{array}{lll} \SetToWidestGAAShift{0,} & \mbox{if} & N_u = N_l, \\ \frac{1}{2} p \sigma_n \gamma_n (N_u - N_l) \cdot & &\\\SetToWidestGAAShift{\cdot (N_l + N_u + 1),}& \mbox{else.}& \end{array}\right.\\
\overline{d_{C}} &=& \SetToWidestGAAShift{p \sigma_n (N_u + 1) d_{Nu}.}\\

\end{array}
\end{equation}
The following probability segments are added to $\Xi_{n+1}$ for $m_{n + 1} = g$: 
\begin{equation}
\begin{array}{l@{\hspace{0.6em}}c@{\hspace{0.1em}}l@{\hspace{0.1em}}}
\SetToWidestGAAShift{-,}& \mbox{if} & d_{Nl} < d_s,\\
\SetToWidestGAAShift{{[T_s - d_{Nl}, T_s - d_s]} \gets p,}&\mbox{else.}& \\
\SetToWidestGAAShift{-,}& \mbox{if} & N_u = N_l,\\{[T_s - \gamma_n, T_s - d_s]} \gets p \cdot (N_u - N_l),& \mbox{else.}\\
\SetToWidestGAAShift{{[T_s - \gamma_n, T_s - \gamma_n + d_{Nu}] \gets p,}}& \mbox{if} & d_{Nu} < \gamma_n - d_s,\\{[T_s - \gamma_n,T_s - d_s]}\gets p,& \mbox{else.}
\end{array}
\end{equation}

If $\gamma_n < d_s$, then no probability is added to the resulting probability buffer and the partial latency contribution added to $\overline{d_{nd}}$ can be computed by 
\begin{equation}
\begin{array}{lll}
\overline{d_p} &=& p \sigma_n \cdot (d_{Nl} N_l \frac{1}{2} \gamma_n (N_u - N_l)  \cdot  {(N_u + N_l + 1)} +\\
&& d_{Nu} (N_u + 1)).
\end{array}
\end{equation}
\\
\textbf{Case 3): $\mathbf{N_u < N_l}$}\\
Like in \textit{growToRight()}, this case has three subcases. If \mbox{$N_l \gamma_n - r \geq d_s$} holds true, the partial latency is (for both $m_{n+1} = s$ and $m_{n+1} = g$)
\begin{equation}
\overline{d_p} = p N_l \sigma_n d_f.\\
\end{equation}
The following segments are added to $\Xi_{n+1}$:
\begin{equation}
\begin{array}{l@{\hspace{1.0em}}l@{\hspace{1.0em}}l@{\hspace{1.0em}}}
{[\gamma_n - d_{Nl}, d_{Nu}]} \gets p, & \mbox{if} & m_{n+1}=s, \\
{[T_s - d_{Nl}, T_s + d_{Nu} - \gamma_n]} \gets p, & \mbox{if} & m_{n+1}=g. \\
\end{array}
\end{equation}
For $ N_l \gamma_n - r < d_s \land d_{Nl} \geq d_s$, it is
\begin{equation}
\overline{d_{p}} = \left\{\begin{array}{lll} p \sigma_n ((d_{Nl} - d_s) (N_l - 1) + & & \\+ N_l(d_s - \gamma_n + d_{Nu})),& \mbox{if} & m_{n+1} = s,\\ p \sigma_n N_l (r - l),& \mbox{if} & m_{n+1} = g.\end{array}\right.
\end{equation}
The following segments are added to $\Xi_{n+1}$:
\begin{equation}
\begin{array}{l@{\hspace{1.0em}}l@{\hspace{1.0em}}l@{\hspace{1.0em}}}
{[\gamma_n - d_{Nl}, \gamma_n - d_s]} \gets p, & \mbox{if} & m_{n+1} = s,\\
{[T_s - d_{Nl}, T_s - d_s]} \gets p, & \mbox{if} & m_{n+1} = g.\\
\end{array}
\end{equation}

If  $N_l \gamma_n - r < d_s \land d_{Nl} < d_s $ holds true or if $\gamma_n \leq d_s$, no probability is added to the resulting buffer and the partial latency is $\overline{d_p} = p \sigma_n N_l d_f$.

\section{Proofs of the Properties}
\label{seq:proofs}
In Section \ref{sec:gammaSequences}, we have defined $\gamma$-sequences and have introduced multiple properties, for which we claimed that they hold true for some $(i_n, j_n)$. In this section, we proof that these properties can be fulfilled.

\subsection{Property \ref{prop:constantShrinkage}: Constant Shrinkage/Growth}
Given an offset $\Phi$ between a scan window and an advertising packet, Property \ref{prop:constantShrinkage} implies that this offset will be by $\gamma_n \geq 0$ time-units larger or smaller for a scan window that is $j_n$ scan intervals and an advertising packet that is $i_n$ advertising intervals later.

Let us consider $i_n \cdot T_a$  and $j_n \cdot T_s$ time-units. Their difference is $|j_n \cdot T_s - i_n \cdot T_a| = \gamma_n$. Let $t_s$ be an arbitrary point in time a scan window starts at, and $t_a$ an arbitrary point in time an advertising packet is sent at, and $\Phi = t_s - t_a$.
Then, for the offset $\Phi'$ that is $i_n$ advertising intervals and $j_n$ scan intervals later, it is 
\begin{equation}
\begin{array}{ccl}
\Phi' &= & t_s + j_n \cdot T_s - (t_a + i_n  \cdot T_a) =\\
& = &t_s - t_a +(j_n \cdot T_s - i_n \cdot T_a)=\\
& = & \Phi \pm \gamma\mbox{ } \forall (t_a, t_s).
\end{array}
\end{equation}
\hfill $\square$
\subsection{Property \ref{prop:rangeOfGamma}: Range of $\gamma$}
Property \ref{prop:rangeOfGamma} implies that there exist pairs $(i_n, j_n)$, such that $0 \leq \gamma_n \leq \min(T_s, T_a)$.
Let us first assume $T_s < T_a$. 
It is $\gamma = |j_n \cdot T_s - i_n \cdot T_a|$. We choose $j_n = \lceil \frac{i_n T_a}{T_s} \rceil$, and it is
\renewcommand\arraystretch{1.5}
\begin{equation}
\begin{array}{ccl}
\gamma_n & = & \left|\left\lceil \frac{i_n T_a}{T_s} \right\rceil T_s - i_n T_a\right|\\
& \leq & \left|\left(\frac{i_n \cdot T_a}{T_s} + 1\right) T_s - i_n \cdot T_a\right|\\
& = & |i_n T_a + T_s - i_n T_a|\\
& = & T_s
\end{array}
\end{equation}

Similarly, if $T_s > T_a$, we can choose $j_n = \lceil\frac{i_n T_s}{T_a}\rceil$, which will result in $\gamma_n \leq T_a$.
\hfill $\square$
\subsection{Property \ref{prop:gammaRecursiveDefinition}: Construction of $\gamma$}
Property \ref{prop:gammaRecursiveDefinition} claims that there are linear combinations 
$\gamma_n = |T_s - |\sum_{k = 0}^{n-1} a_k \cdot \gamma_{k}||$, $a_k \in \mathbb{Z} \setminus 0$, which fulfill the following three properties.
\begin{itemize}
	\item $\gamma_{n} \leq \frac{1}{2} \gamma_{n-1}$
	\item the sum $\sum_{k=0}^{n-1} |a_k| \cdot (i_k T_a)$ is minimized, and 
	\item for a shrinking sequence, $a_k < 0$, otherwise, $a_k > 0$
\end{itemize}
A valid constructive algorithm for linear combinations that fulfill these properties has been presented in Section \ref{sec:algo}, which proofs the existence of such linear combinations.
\hfill $\square$

\subsection{Property \ref{prop:lowestOrderFirst}: Priority of Lower Orders}
This property claims that for any $\gamma_n$, $\sigma_n = i_n \cdot T_a$ is always at least as large as the smallest possible linear combination $\sum_{k=0}^{n-1} |a_k| \cdot \sigma_k$, for which $|\sum_{k=0}^{n-1} a_k \gamma_k| \geq T_s - \gamma_n \geq T_s - \frac{1}{2} \gamma_{n-1}$.
The construction of $\gamma_n$ is defined by $\gamma_{n} = | T_s - |\sum_{k = 0}^{n-1} a_k \cdot \gamma_{k}||$, $a_k \in \mathbb{Z}  \setminus 0$, such that i) $\gamma_{n} < \frac{1}{2}\gamma_{n-1}$ and ii) the corresponding penalty $\sigma_n$ is minimized. The necessary condition $\gamma_n = |T_s - |\sum_{k = 0}^{n-1} a_k \cdot \gamma_{k}|| < \frac{1}{2} \gamma_{n-1}$ implies that $|\sum_{k = 0}^{n-1} a_k \cdot \gamma_{k}| \geq T_s - \frac{1}{2} \cdot \gamma_{n-1}$. This sum has a corresponding sum of penalties $\sum_{k=0}^{n-1} |a_k| \cdot \sigma_k$. Since $\gamma_n = |T_s - |\sum_{k = 0}^{n-1} a_k \cdot \gamma_{k}||$, it is $|\sum_{k = 0}^{n-1} a_k \cdot \gamma_{k}| \geq T_s - \gamma_n$.
\hfill $\square$

\subsection{Property \ref{prop:limitedOrder}: Highest Possible Order}
\label{sec:maxIterationProof}
The maximum number of iterations of Algorithm~\ref{alg:algorithmDefintion} is identical to the maximum order $n_m$. In the following, we proof that the maximum order is bounded, as described by Property~\ref{prop:limitedOrder}.

Property~\ref{prop:gammaRecursiveDefinition} implies that $\gamma_{n} \leq \frac{1}{2}\gamma_{n-1}$. Hence, in the worst case, $\gamma_{n} = \frac{1}{2}\gamma_{n-1}\mbox{ } \forall n$.
Using this, we can define a worst-case $\gamma$-parameter for the order $n$, $\gamma_{wc,n}$ by assuming a maximum initial value $\gamma_0$ of $\min(T_a, T_s)$ (cf. Property \ref{prop:rangeOfGamma}). It is:
\begin{equation}
\gamma_{n,wc} = \left(\frac{1}{2}\right)^{n} \cdot \min(T_a, T_s).
\end{equation}
In the worst-case, the highest-order sequence is the first one for which \mbox{$\gamma_{n,wc} \leq d_s$} holds true. Hence, it is
\begin{equation}
\frac{1}{2}^{n_m} \cdot \min(T_a, T_s) \leq d_s.
\end{equation}
Solving this inequality leads to a maximum order $n_m$, as given by Equation \eqref{eq:maxIterationOrderProof}.
\begin{equation}
\label{eq:maxIterationOrderProof}
n_m = \left\lceil \frac{\ln(\min(T_a,T_s))-\ln(d_s)}{\ln(2)} \right\rceil
\end{equation}
In addition, it is intuitively clear that if $\min(T_a, T_s) \leq d_s$, only order-0-sequences can occur (the advertiser can never ``overtake'' the scanner) and therefore, $n_m = 0$. Further, we have assumed $d_s$ to be shortened by $d_a$ time-units, as described in Section \ref{sec:overview}, and the equation above is only valid for shortened values of $d_s$. 

The limited number of sequence orders also implies that whenever $\gamma_n \ne 0 \mbox{ } \forall n \leq n_m$, the worst-case discovery latency is bounded. This is because the discovery latency is defined by the sum of $\gamma$-intervals that fit into the scan interval, which is always finite if $\gamma \ne 0$.

\hfill $\square$
\subsection{Proof of Latency Computation}
\label{sec:latencyCompProof}
In Section \ref{sec:algo}, we form linear combinations ${L' = \sum_{k=0}^n a_k \cdot \gamma_k}$, such that a packet that is sent with a certain initial offset $\Phi[0]$ from the first scan window is shifted by $L'$ towards the next scan window. We have claimed that coefficients $a_k$ for high values of $k$ always need to be minimized, no matter how by how much coefficients $a_k$ for lower values of $k$ need to be increased to compensate for this minimization. In what follows, we proof this claim.

Let us consider two linear combinations $L'_1$ and $L'_2$ that both fulfill $- d_s \leq \Phi[0] + L' \leq 0$, or $T_s - d_s \leq \Phi[0] + L' \leq T_s$, respectively, and hence lead to a coincidence of the advertising packet and a scan window. Further, $L'_1 = \sum_{k=0}^{n} a_{k,1} \cdot \gamma_k$, whereas $L'_2 = \sum_{k=0}^{n-1} a_{k,2} \cdot  \gamma_k$. Further, $a_{k,1}$ and $a_{k,2}$ are chosen such that the corresponding sum of penalties is minimized. We now study whether $L'_1$ or $L'_2$ has a lower sum of penalties.

If $m_{n-1} = m_{n}$, then $\sum_{k=0}^{n-1} a_{k,2} \cdot \gamma_k - \sum_{k=0}^{n-1} a_{k,1} \cdot  \gamma_k \leq \gamma_{n-1}$, since $\gamma_{n} \leq \frac{1}{2} \gamma_{n-1}$. This means that by having $1 \cdot \gamma_n$ in $L'_1$, at most one instance of $\gamma_{n-1}$
can be saved. Because $\sigma_n \geq \sigma_{n-1}$, $L'_2$ always has a lower sum of penalties than $L'_1$.

\begin{figure}[htb]
	\centering
	\includegraphics[width = 1.0\columnwidth]{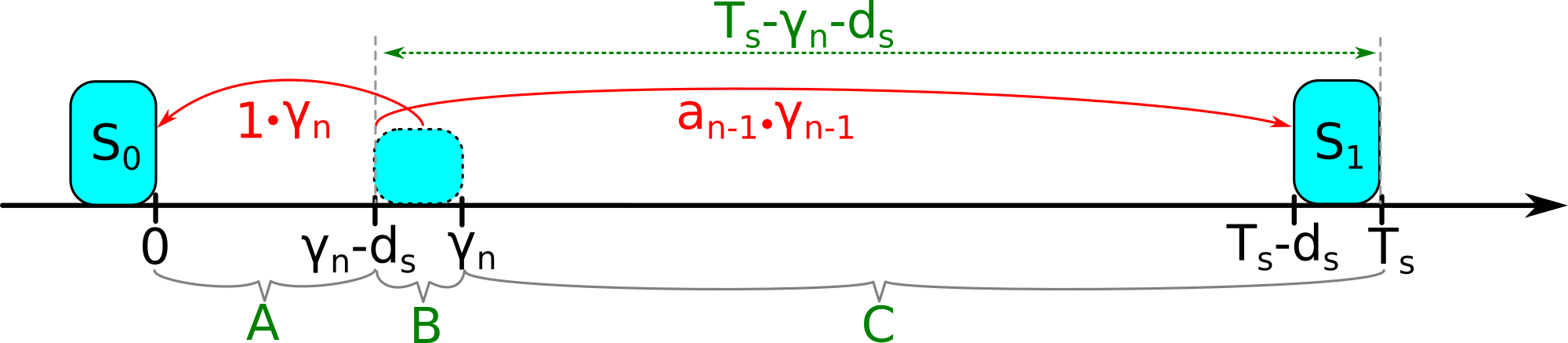}
	\vspace*{-0.5cm}
	\caption{Proof for Latency Computation}
	\label{fig:proof_partitioning} 
\end{figure}

If $m_{n-1} \ne m_n$, the initial offset $\Phi[0]$ could potentially form a situation in which $L'_1$ is defined by $1 \cdot \gamma_{n}$, and ${L'_2 = \sum_{k=0}^{n-1} a_k \gamma_k \geq T_s - d_s}$ in the worst-case. In other words, there might be situations in which either $1 \cdot \gamma_n$, or an appropriate number of $\gamma_{n-1}$ time-units of length $T_s - d_s$ or longer, will lead to a match. This can be seen as the worst case. Such a situation with $\gamma_{n-1} = g$ and $\gamma_n = s$ is depicted in Figure~\ref{fig:proof_partitioning}. We now study which of both possibilities $L'_1, L'_2$ has a lower sum of corresponding penalties. 

Here, the range of initial offsets $\Phi[0]$ between $\gamma_n - d_s$ and $\gamma_n$ (Area \textbf{B} in Figure \ref{fig:proof_partitioning}) forms the only case in which the left scan window $S_0$ can be reached by $1 \cdot \gamma_n$. 
From Property~\ref{prop:lowestOrderFirst} follows that the corresponding penalty $1 \cdot \sigma_n$ will be at least the lowest possible sum of penalties that corresponds to a linear combination $|\sum_{k=0}^{n-1} a_k \cdot \gamma_k| \geq T_s - \gamma_n$. However, as can be seen from the figure, the temporal distance from the right scan window $S_1$ of the entire area \textbf{B} is no more than $T_s - d_s - \gamma_n < T_s - \gamma_n$ time-units. Hence, if there is a linear combination such that $\Phi[0] + \sum_{k=0}^{n-1} a_k \cdot \gamma_k$ lies within $S_1$, the penalty of $\sigma_n$ will be at least as high as the sum of penalties $\sum_{k=0}^{n-1} a_k \cdot \sigma_k$. For the same reason, in area \textbf{C}, the latency of a suitable linear combination of $\sum_{k=0}^{n-1} a_k \cdot \gamma_k$ will be always lower than any linear combination that contains at least $1 \cdot \gamma_n$. 
Initial offsets $\Phi[0] < \gamma_n - d_s$ in area \textbf{A} can either be formed by $a_n \cdot \gamma_n + \sum_{k=0}^{n-1} a_k \gamma_n$, with $a_n \geq 1$ and $a_k > 0$ for at least some $k$, or by linear combinations without $\gamma_n$, i.e., $\sum_{k=0}^{n-1} a_k \cdot \gamma_k$. Clearly, $\sum_{k=0}^{n-1} a_k \cdot \gamma_k$ is the possibility with the lowest sum of penalties.
\hfill $\square$

\section{Relation to the Euclidean Algorithm}
Property \ref{prop:gammaRecursiveDefinition} defines $\gamma_n$ as $\gamma_{n} = | T_s - L_{\gamma_{n-1}}||$, where $L_{\gamma_{n-1}}$ is a linear combination $\sum_{k = 0}^{n-1} a_k \cdot \gamma_{k}$. Let us consider two possibilities for $L_{\gamma_{n-1}}$, both assuming $m_0 = g$. The one, $L_s$, is the linear combination $\sum_{k=0}^{n-1} a_k \gamma_k$ that approximates $T_s$ as close as possible, without exceeding it. The other one, $L_g = L_1 + \gamma_{n-1}$ slightly exceeds $T_s$, and $\min({|T_s - L_1|},\mbox{ } {|T_s - L_2|}) < \frac{1}{2}\gamma_{n-1}$. If $L_s < L_g$, $m_n = s$ and $\gamma_{n} = T_s - L_s$. Otherwise, $m_n = g$ and $\gamma_{n} = L_g - T_s$. This shares similarities with the method of least absolute remainders, which the mathematician Leopold Kronecker has shown to be the most efficient version of the Euclidean algorithm \cite{ore:48}. For determining the next remainder (here: $|T_s - L_{\gamma_{n-1}}|$), rather than dividing the smallest previous reminder by its predecessor (as in the Euclidean algorithm), we divide the larger previous reminder (i.e., $d_{t,n-1}$) by the smaller one (i.e. $\gamma_{n-1}$.).

\section{Definition of the Latency Computation Algorithm}
The algorithm described in Section \ref{sec:algo} is formally defined by Algorithm \ref{alg:algorithmDefintion}.
\label{app:algorithm}
\begin{algorithm}[h!]
	\caption{High-level algorithm for computing $d_{nd}$}
	\label{alg:algorithmDefintion}
	\small
	\begin{algorithmic}[1]
		\Require $T_a\mbox{, }d_a\mbox{, }T_s\mbox{, }d_s$
		\Ensure $\overline{d_{nd}}$
		%\If {$T_a \leq d_s$}													  \label{al:checkNongamma}
		%\State	$\overline{d_{nd}} \gets nonGammaMatching(T_a, T_s, d_s) + d_{a}$ \label{al:nonGammaMatching}
		%\Else
		\State $(\overline{d_{nd}}, \Xi_{0}, \vec{\gamma}, \vec{\sigma}, \vec{m}) \gets Initialize(T_a, T_s, d_s)$  \label{al:initalize}
		\If {$T_a \leq T_s$}													  
		\State $n \gets 0$
		\Else
		\State $n \gets 1$
		\EndIf
		\While {$\gamma_n \geq d_s$} \label{al:loopBegin}
		\If{$m_n = g$} \label{al:checkMode}
		\State $(\overline{d_{nd}}, \Xi_{n+1}) \gets growToRight(\overline{d_{nd}}, \Xi_{n},\gamma_n, \sigma_n, m_{n+1})$ \label{al:gammarizeGrowing}
		\ElsIf{$m_n = s$}
		\State $(\overline{d_{nd}}, \Xi_{n+1}) \gets shrinkToLeft(\overline{d_{nd}}, \Xi_{n},\gamma_n, \sigma_n, m_{n+1})$ \label{al:gammarizeShrinking}
		\Else
		\State $\overline{d_{nd}} \gets \infty$; return
		\EndIf
		\If {$(\|\Xi_{n} \| = \emptyset) \lor (\gamma_n < d_s$)} 
		\State break loop
		\Else
		\State $n \gets n+1$	\label{al:npp}
		\EndIf
		\EndWhile
		\State $\overline{d_{nd}} = \overline{d_{nd}} + d_a$  \label{al:pkgLength}
		%\EndIf
	\end{algorithmic}
\end{algorithm}
\pagebreak
\section{Software Library}
\label{sec:using_the_model}
Our proposed model can be downloaded as a MATLAB library from \url{www.rcs.ei.tum.de/wireless_systems}. 
The function 
\begin{verbatim}
function [dAvg, dMin, dMax, o] = 
getDiscoveryLatency(Ta,Ts,dA,ds,iL)
\end{verbatim}
computes the mean- (\textit{dAvg}), minimum- (\textit{dMin}) and maximum- (\textit{dMax}) discovery latency for a given parametrization. The parameters of this function are the advertising interval (\textit{Ta}), the scan interval (\textit{Ts}), the scan window (\textit{ds}), the advertising packet-length (\textit{dA}) and  an iteration limit (\textit{il}) of the algorithm. The algorithm also returns the maximum sequence order (\textit{o}) for the given parameters. An example for computing entire latency curves with sweeping parameters is available for download, too.
\section{Table of Symbols}
\small
\label{sec:table_of_symbols}
\renewcommand{\arraystretch}{1.10}
\begin{tabularx} {\columnwidth}{p{0.75cm}|X}
	$T_a$ & Advertising Interval\\ 
	$T_s$ & Scan Interval \\ 
	$d_s$ & Scan Window \\ 
	$d_a$ & Duration of an advertising packet\\
	$k$ & Counting index\\
	$n$ & Index for the order of a sequence\\
	$n_m$ & Maximum sequence order \\
	$h,i,j$ & Generic indices / integer multiples\\
	$\mathbb{N}_0$ & Set of all non-negative integers\\
	$\Phi$ & Offset between an advertising packet and the closest neighboring scan window on its left side\\
	$\Phi'$ & Offset between an advertising packet and the closest neighboring scan window on its right side\\
	$d_{nd}$ & Neighbor discovery latency \\
	$\overline{d_{nd}}$& Mean neighbor discovery latency \\
	$d_{nd,m}$ & Max. neighbor discovery latency \\
	$\gamma$ & Growth/Shrinkage per interval \\
	$\gamma_s$,$\gamma_g$ &$\gamma$-parameter given the sequence is shrinking/growing\\
	$\gamma_{wc}$ & Worst-case value of $\gamma$\\
	$N_i$ & Number of intervals until reaching the next target area\\
	%$X_n$ & Number of advertising intervals to shrink a certain offset\\
	$m$ & Mode of a sequence (s $\gets$ shrinking, g $\gets$ growing, \mbox{c $\gets$ coupling})\\
	$d_t$ & Distance left to travel\\
	$\sigma$, $\sigma_s$&Penalty related to $\gamma_n$ / Penalty related to $d_{t,n}$ \\
	$i_n,\mbox{ }j_n$ & Multiplier of advertising ($i_n$) and scan intervals ($j_n$)\\
	$L_{\gamma_n}$ & Linear combination $\sum_{k=0}^{n} a_k \cdot \gamma_k$\\
	$a_k$ & Coefficient of a linear combination\\
	%$\sigma_s$ &Penalty sum\\
	$\Xi$ & Probability buffer\\
	$\| \Xi \|$ & Number of segments in the probability buffer $\Xi$\\
	$\Xi[k]$ & Probability density of segment $k \in \Xi$ \\
	$t_s[k]$ & Starting time of a segment $k \in \Xi$\\
	$t_e[k]$ & Ending time of a segment $k \in \Xi$\\
	$l$,$r$ & Abbreviations for $t_s[k]$ and $t_e[k]$\\
	$d_{lo}$ & Leftover duration (Duration of a partition which does not fit entirely in the partition of its previous-order sequence)\\
	$d_{loh}$ & Hitting duration of a leftover partition\\
	$d_{lom}$ & Missing duration of a leftover partition\\
	%$p$ & Probability \\
	$\overline{d_{p}}$ & Weighted partial duration of a partition \\
	$p_i$, $d_i$ & Probability/Latency of a segment $i$ of a probability buffer\\
	$N_{sh}$ & Number of intervals to shrink a certain distance\\
	$t$/$t'$ & Original/Transformed point in time \\
	$N_l$, $N_u$ & Minimum/Maximum number of $\gamma$-intervals until reaching a target area for a segment of $\Xi$\\
	$d_{Nl}$, $d_f$, $d_{Nu}$ & Length of the part of a segment that has ${N_l-1}$/${N_l...N_u}$/${N_u + 1}$ intervals to travel until reaching a target area\\
	%$d_A$ $d_B$ $d_C$ & Partial latencies of part A/B/C of a segment\\
	$\zeta[k]$ & Largest sum of penalties for a segment $k \in \Xi$\\
	$d_{p,m}$ & Maximum partial latency for a part of a segment\\
	$\O$, $\O_m$ & Root mean square error (RMSE) for mean/maximum latencies\\
	$\kappa$, $\kappa_m$ & Normalized root mean square error for mean/maximum latencies\\
	$MD$, $MD_m$ & Maximum deviation ($\max(\vert d_{comp} - d_{sim} \vert)$) for mean/maximum latencies\\
	$d_{comp}$, $d_{sim}$ & Computed/Simulated latency\\
\end{tabularx} 
\end{appendix}
\end{document}